% ...
\documentclass[useAMS,usenatbib]{mn2e}

\usepackage{color}
\usepackage{graphicx}
\usepackage{lscape}
\usepackage{amsmath}
\usepackage{amssymb}
\usepackage{rotating}

\newcommand\aj{\rmfamily{AJ}}          % Astronomical Journal
\newcommand\apj{\rmfamily{ApJ}}        % Astrophysical Journal
\newcommand\apjs{\rmfamily{ApJS}}      % Astrophysical Journal, Supplement
\newcommand\aaps{\rmfamily{A\&AS}}     % Astronomy and Astrophysics, Supplement
\newcommand\aap{\rmfamily{A\&A}}       % Astronomy and Astrophysics
\newcommand\mnras{\rmfamily{MNRAS}}    % Monthly Notices of the RAS
\title[Testing spectral models for stellar populations]{Testing spectral models for stellar populations with star clusters: II. Results}
\author[Rosa M. Gonz\'alez Delgado and Roberto Cid Fernandes]{Rosa M. Gonz\'alez Delgado$^{1}$\thanks{E-mail:
rosa@iaa.es} and Roberto Cid Fernandes$^{2}$\thanks{E-mail:cid@astro.ufsc.br}\\
$^{1}$Instituto de Astrof\'\i sica de Andaluc\'\i a (CSIC), P.O. Box 3004, 18080 GRANADA, Spain \\
$^{2}$Departamento de F\'{\i}sica-CFM, Universidade Federal de Santa Catarina, C.P.. 476, 88040-900, Florian\'opolis, SC, Brazil}

\begin{document}

\date{2009 July}

%pagerange{\pageref{firstpage}--\pageref{lastpage}} \pubyear{2008}

\maketitle

\label{firstpage}

\begin{abstract}
High spectral resolution evolutionary synthesis models have become a
routinely used ingredient in extragalactic work, and as such deserve
thorough testing. Star clusters are ideal laboratories for such
tests. This paper applies the spectral fitting methodology outlined in
Paper I to a sample of clusters, mainly from the Magellanic Clouds and
spanning a wide range in age and metallicity, fitting their integrated
light spectra with a suite of modern evolutionary synthesis models for
single stellar population. The combinations of model plus spectral
library employed in this investigation are Galaxev/STELIB,
Vazdekis/MILES, SED@/GRANADA, and Galaxev/MILES+GRANADA, which provide
a representative sample of models currently available for spectral
fitting work. A series of empirical tests are performed with these
models, comparing the quality of the spectral fits and the values of
age, metallicity and extinction obtained with each of them. A
comparison is also made between the properties derived from these
spectral fits and literature data on these nearby, well studied
clusters.  These comparisons are done with the general goal of
providing useful feedback for model makers, as well as guidance to the
users of such models. We find that: (1) All models are able to derive
ages that are in good agreement both with each other and with
literature data, although ages derived from spectral fits are on
average slightly older than those based on the S-CMD method as
calibrated by Girardi et al.\ (1995). (2) There is less agreement
between the models for the metallicity and extinction. In particular,
Galaxev/STELIB models underestimate the metallicity by $\sim$0.6 dex,
and the extinction is overestimated by 0.1 mag. (3) New generation of
models using the GRANADA and MILES libraries are superior to
STELIB-based models both in terms of spectral fit quality and
regarding the accuracy with which age and metallicity are retrieved.
Accuracies of about 0.1 dex in age and 0.3 dex in metallicity can be
achieved as long as the models are not extrapolated beyond their
expected range of validity.
\end{abstract}


\section{Introduction}
\label{sec:Introduction}

Evolutionary synthesis has progressed significantly since this
technique was introduced by Tinsley (1968). In particular,
high/intermediate spectral resolution stellar libraries like STELIB
(Le Borgne et al.\ 2003), INDO-US (Valdes et al.\ 2004), ELODIE
(Prugniel \& Soubiran 2004), GRANADA (Martins et al.\ 2005); INAOE
(Rodr\'\i guez-Merino et al.\ 2005), and MILES (S\'anchez-Bl\'azquez
et al.\ 2006) have been incorporated into evolutionary synthesis codes
(e.g.\ Bruzual \& Charlot 2003, hereafter BC03; Le Borgne et al.\
2004; Gonz\'alez Delgado et al.\ 2005) whose predictions are now
routinely used in the analysis of galaxy spectra.  With the
proliferation of evolutionary synthesis models, it becomes critically
important to test them, specially in light of the fact that a enormous
fraction of extragalactic studies are heavily dependent on them.

Because of their complex mixture of stellar populations, galaxies are
not ideal test beds for evolutionary synthesis models.  Star clusters
(SCs), on the other hand, are suitable laboratories for this
purpose. Presumably formed in a single burst, SCs can be characterized
by a single age ($t$), a single metallicity ($Z$), and an extinction
($A_V$). Even taking into account caveats like stochastic effects
(Cervi\~no et al.\ 2002; Ma\'\i z-Apell\'aniz 2009) and the
possibility of non-unique populations (e.g. NGC 1569, Gonz\'alez
Delgado et al.\ 1997; and $\omega$ Cen, Meylan 2003), SCs are by far
and large the simplest systems available to test evolutionary
synthesis models.  Furthermore, SC properties can be determined from
spatially resolved observations like color-magnitude diagrams (CMD)
and spectroscopy of individual stars, providing a benchmark for
empirical tests of integrated light models.  SC integrated light,
through colors or spectra, have also been very useful to test stellar
population models, in particular the clusters in the Magellanic Clouds
(e.g. Beasley et al.\ 2002; de Grijs \& Anders 2006; Pessev et al.\
2008; Santos et al. 2006).

Studies in this general vein have been carried out in the last couple
of years.  Wolf et al (2007) fitted low resolution spectra of SCs
using the BC03 models, finding ages and metallicities in good
agreement with independent estimates for ages $> 1$ Gyr. For younger
clusters the agreement in ages is still acceptable (0.3 dex), but
metallicities cannot be well constrained. More recently, Koleva et
al.\ (2008) have fitted high resolution spectra of SCs with STELIB,
MILES and ELODIE based models. Results for the latter two libraries
are consistent with each other and with ages and metallicities derived
from spatially resolved studies, while the BC03 STELIB based models
produce discrepancies associated with the incompleteness of this
library, specially in metallicity. Their results validate full
spectral fitting as a means to estimate SC properties, while at the
same time highlighting the importance of stellar libraries in this
game.

This paper follows this same general direction. We present results of
a $\lambda$-by-$\lambda$ spectral synthesis analysis of SCs obtained
with different sets of high resolution evolutionary synthesis
models. The main differences between this study and that by Wolf et
al.\ (2007) are that they focus on only one set of BC03 models, and
use data covering a wider wavelength interval but at coarser spectral
resolution than the one used here. Compared to Koleva et al.\ (2008),
our study differs mostly in the properties of the target SCs. Our
sample is composed mainly of young and intermediate age SCs in the
Large and Small Magellanic Clouds (LMC and SMC, respectively), whereas
Koleva et al.\ concentrate on an older population of Galactic globular
clusters. There also differences in the fitting methodology, like the
fact that our fits do take into account the continuum shape, whereas
these two studies apply spectral rectification techniques.  Together,
these complementary approaches provide valuable feedback to
evolutionary synthesis modelers, as well as guidance to users of these
models.

This study started in Paper I (Cid Fernandes \& Gonz\'alez Delgado
2009), where we presented our spectral fitting methodology. High
quality spectra in the 3650--4600 \AA\ range of 27 SCs from the sample
of Leonardi \& Rose (2003, hereafter LR03), were fitted with single
stellar population (SSP) spectral models from Vazdekis et al.\ (2009),
based on the MILES library. The fits were carried out with the
publicly available {\sc starlight} code (Cid Fernandes et al. 2005,
2008), never before used for SC work. Covariance maps and a simple
bayesian parameter estimation formalism were also presented.  In this
second paper, the same methodology is extended to other publicly
available high resolution evolutionary models.  The results are
compared both among themselves and to independent estimates of $t$ and
$Z$. This allows us to (1) compare the quality of spectral fits
obtained with different models, (2) quantify uncertainties and biases
in SC properties resulting from differences between different models,
and (3) check the reliability of the ages and metallicities derived
from detailed spectral fitting as compared to more traditional methods
\footnote{We are not able to derive the photometric mass of each
cluster because, even though the spectra are flux calibrated, the
absolute scale is unknown.}.

This study can be looked at from two different perspectives: (a)
Validation of spectral fitting as a means of inferring the properties
of SCs, or (b) validation of evolutionary synthesis models for use in
extragalactic work. Our main interest is in the latter, and this
influences the way we present our results, but researchers more
interested in SCs per se can also benefit from this study.
 
The SC sample and information about the age, metallicity, and
extinction previously reported in the literature are described in
Section \ref{sec:Sample}. The seven SSP models used here come from
four sources: BC03, Charlot \&Bruzual (2009, in preparation),
Gonz\'alez Delgado et al (2005), and Vazdekis et al.\ (2009). 
Section \ref{sec:STPop_Models} presents a brief
description of each of these. Results start in Section
\ref{ref:SpectralFits}, where we compare the quality of spectral fits
obtained with different models. Section \ref{sec:Results_ModXMod}
compares SC properties derived with different models, whereas in
Section \ref{sec:Synthesis_X_Literature} we compare the results
obtained with spectral synthesis to those previously obtained from
CMD, broad band colors, or spectral indices. The discussion is
presented in Section \ref{sec:Discussion}.  Section
\ref{sec:Conclusions} summarizes our main conclusions.

\section{The sample and literature data}
\label{sec:Sample}

The sample analyzed here comprises the same 27 SCs from LR03 studied
in Paper I, 20 of which are from the LMC, while 3 belong to the SMC
and 4 are Galactic clusters. The spectra cover the 3650--4600 \AA\
wavelength range with a typical $S/N$ of about 50. This spectral range
is very suitable for testing models at young and intermediate ages
because it covers the Balmer jump and the high-order Balmer series
(Gonz\'alez Delgado et al.\ 1999).

The LMC SCs which dominate this sample were originally selected by
LR03 from Sagar \& Pandey (1989), who compiled the ages and
metallicities derived from CMDs in the literature. These SCs cover the
well known range of ages and metallicities in the LMC, reflecting the
age-metallicity gap in that galaxy (e.g. Girardi et al.\ 1995). Thus,
most of them are younger than 2 Gyr (age interval is between 30 Myr
and 2 Gyr), with metallicities over 0.25 solar, while a few are old
and very metal poor. To extend the analysis to the low metallicity and
old age regime, 3 SCs from the SMC (NGC 411, NGC 416 and NGC 419) and
4 from the Galaxy (47 Tuc, M15, M79 and NGC 1851) are included in the
sample. These extra SCs approximately cover the $t$-$Z$ gap of the LMC
SCs.

Fiducial reference values of $t$ and $Z$ determined from fundamental
methods are required for comparison with our spectral fitting
estimates\footnote{We will refer to our results as ``{\sc starlight}''
or ``spectral fitting'' ages, metallicities, and extinctions.}.  CMD
based ages should be the best option. However, due to the lack of
consistency and homogeneity of the results based on the CMDs, we
adopted the values tabulated by LR03.

LR03 adopted ages which are derived from a secondary indicator, the
$S$ parameter, based on $U-B$ and $B-V$ integrated colors. The $S$
parameter provides an empirical relation between the age of a SC and
its integrated colors (Elson \& Fall 1989).  LR03 use UBV photometry
from Bica et al.\ (1992, 1996) and the $S$ parameter calibration given
by Girardi et al.\ (1995), based on 24 LMC clusters whose ages are
estimated from high-quality ground-based CMDs. Note, however, that
this relation between $S$ and $\log t$ has a rms dispersion of 0.137
dex. For metallicities we have also adopted the value given in LR03,
taken from Olszewski et al.\ (1991) and based on the equivalent width
of the CaII triplet at 8500 \AA\ as measured in one or several stars
of each SC, and from the compilations from Seggewiss \& Richtler
(1989), and Sagar \& Pandey (1989). These ages and metallicities are
listed in columns 3 and 4 of Table \ref{tab:Data}. We will take these
values as reference for the results of our spectral fitting estimates
of $t$ and $Z$.

Table \ref{tab:Data} also lists the ages and metallicities derived by
LR03 using the Balmer discontinuity, and two absorption indices based
on H$\delta$/FeI$\lambda$4045, and CaII H+H$\epsilon$/CaII K line
ratios.  LR03 found a good correlation between their results and those
from the Girardi et al.\ (1995) calibration, but their ages are about
50$\%$ older. The correlation between the index-based metallicities of
LR03 and those from the CaII triplet are worse, presenting a large
scatter as well as a systematic offset, with the L03 metallicities
systematically smaller than those of Olszewski et al.\ (1991).  Table
\ref{tab:Data} also lists ages and metallicities from CMD analysis,
taken from a variety of sources.  For the four Galactic clusters of
this sample, we have taken the data from the work of Koleva et al.\
(2008) derived from full spectral fitting, but Table \ref{tab:Data}
also quotes the metallicities compiled by Schiavon et al.\ (2005) for
the same SCs. The comparison of these two sets of values provides a
rough estimation of the uncertainties on the literature metallicity
for these four old clusters.

Finally, we have compiled the V-band extinction $A_V$ from the
literature.  Bica \& Alloin (1986) give a global extinction towards
the LMC and SMC that is $A_V \leq 0.2$ and 0.1, respectively.
McLaughling \& van der Marel (2005) use the BC03 models and optical
colors to derive the V-band mass to light ratios and reddening. The
$A_V$ values obtained are also low, ranging between 0.03 and 0.09,
with a mean 0.06 for the SCs of this sample. Pessev et al.\ (2008)
gives the extinction obtained using the web tool by the Magellanic
Clouds Photometric Survey (MCPS, Zaritsky, Harris \& Thompson 1997;
Zaritsky et al.\ 2004), which gives $A_V$ estimates along the
line-of-sight to stars within a search radius of the SC coordinates.
These estimates range between $A_V = 0.17$ and 0.76, with a mean value
of 0.35. This method can only provide a rough estimate of the
extinction.  Pessev et al.\ (2008) also compile the extinction for 9
of the SCs of the sample estimated from CMDs, $A_V$ ranges between
0.03 and 0.37.  The last columns of Table \ref{tab:Data} summarize
these estimates.
 
%The $A_V$ values tabulated by McLaughlin \& van de
%Marel (2005) and Pessev et al.\ (2008) are listed in Table
%\ref{tab:Data}.
 
%*************************************************************************************************
 \begin{table*}
  \centering
   \begin{tabular}{@{}lccccccccccc@{}}
   \hline
Name  & ID   & log age (yr)& [Fe/H]  & log age (yr) & [Fe/H] & log age (yr) & [Fe/H]  &  Ref. & A$_V$ & A$_V$ &  A$_V$\\
            &    &  S                 &              & Indices            & Indices  & CMD     & CMD          &           & CMD  & MCPS    & McL\\
    (1)   & (2)&  (3)             & (4)       & (5)                    & (6)          & (7)         & (8)              &  (9)     & (10)    & (11)        & (12) \\
  \hline
NGC411   & 1  & 9.25 & -0.84 & 9.09 & -0.43 &          &               &             & 0.37 & 0.17 &  \\
NGC416   & 2  & 9.84 & -1.44 & 9.82 & -1.27 &          &               &              & 0.25 & 0.20 & \\
NGC419   & 3  & 9.08 & -0.70 & 9.22 & -0.90 &           &             &               &          & 0.20 & \\
NGC1651 &4   &9.20 & -0.37 & 9.34 & -0.82  &  9.30 & -0.70 & Ke07     &0.34 & 0.35 & 0.03  \\
                    &     &         &           &          &            &   9.30 &          & Ge97     &         &          &     \\
NGC1754  & 5 & 10.2 & -1.42 & 9.74 & -1.44 &  10.19&        &  Ol98      & 0.28 & 0.40& 0.05   \\
NGC1783  & 6 & 8.94 & -0.75 & 9.19 & -0.54 & 9.11   &          & Ge97    &        & 0.30 &     \\
NGC1795 & 7  & 9.23 & -0.23 & 9.30 & -0.69 & 9.23   &           & Ge97   &        &           &  \\
NGC1806 & 8  & 8.70 & -0.23 & 9.28 & -0.64 &           &             &             &       & 0.25  &    \\
NGC1818 & 9  & 7.54 & -0.90 & 7.30 &           &           &              &             &       & 0.39 & 0.05\\
NGC1831 & 10 & 8.57 & 0.010& 8.70 & -0.65& 8.60 & 0.01     & Gi95   &0.34& 0.39 & 0.04 \\
                   &       &          &            &          &          & 8.85 & -0.01   & Ke07   &       &          &      \\
NGC1846 & 11 & 9.09 & -0.70 & 9.50 & -1.40&          &              &             &       & 0.41 &     \\
NGC1866 & 12 & 8.34 & -1.20 & 8.48 &          & 8.14  & -0.4:    & Gi95    &       & 0.28 & 0.03\\
NGC1978 & 13 & 9.23 & -0.42 & 9.41 & -0.72&  9.3   &             & Ge97  &        & 0.76 &     \\
NGC2010 & 14 & 8.20 &  0.00 & 7.92 &           & 8.19  & -0.4:   &  Gi95   &        &         &   \\
NGC2133 & 15 & 8.11 & -1.00 & 8.16 &          &            &            &             &        &         &       \\
NGC2134 & 16 & 8.28 & -1.00 & 8.80 &          & 8.15   & -0.4:   & Gi95   &        & 0.62 &  \\
NGC2136 & 17 & 8.04 & -0.40 & 7.91 &          &  8.00  & -0.55 & Di00    &        & 0.58  & 0.09\\
NGC2203 & 18 & 9.26 & -0.52 & 9.19 & -0.46& 9.25   &           & Ge97   &        & 0.39 &   \\
NGC2210 & 19 & 10.1 & -1.97 & 9.58 & -1.16& 10.2   &           &  Ge97  & 0.28& 0.39 & 0.03\\
NGC2213 & 20 & 9.01 & -0.01 & 9.32 & -0.88& 9.23   & -0.70 & Ke07  & 0.19 & 0.40 & 0.08\\
                   &        &         &           &           &         &   9.20  &           & Ge97  &          &         &    \\
NGC2214 & 21  & 7.91& -1.20 & 7.72 &          & 7.92   & -0.4:  & Gi95    &           & 0.39 & 0.09 \\
NGC2249 & 22  & 8.72& -0.05 & 8.44 & -0.40& 8.54   & -0.4   & Gi95   &  0.03 & 0.39 & 0.06\\
                   &        &         &            &         &           &  9.0    & -0.45 & Ke07 &            &         &       \\
NGC2121 & 23  &         &            &         &           &  9.46  & -0.40 & Ke07 & 0.22   & 0.53 & 0.10 \\
                    &       &         &            &          &          & 9.3     &           & Ge97 &           &           &     \\
47Tuc        &  24 &         &            &10.1  &-0.76, -0.70* &            &           &            &           &           &  \\
M15           & 25  &         &            & 10.0 & -2.29&            &           &            &           &            & \\
M79           & 26  &         &            & 10.1 &-1.95, -1.55*&            &           &            &           &            &   \\
NGC1851 & 27  &        &             & 9.72 &-1.16, -1.21* &           &            &            &          &            &         \\
 \hline
 \end{tabular}
\caption{Properties (age, metallicity and extinction) of the SCs from
the literature.  Col.\ (1): SC name.  Col.\ (2): SC ID number.  Col.\
(3): Ages derived using the S-parameter from Girardi et al.\ (1995)
and tabulated by LR03.  Col.\ (4): Metallicity obtained with the
equivalent width of the CaII triplet from Olszewski et al.\ (1991),
and compilation by Seggewiss \& Richtler (1989), and Sagar \& Pandey
(1989).  Col.\ (5) and (6): Ages and metallicities obtained by LR03
using the CaII and H$\delta$/Fe indices for the SMC and LMC SCs. The
data for the four Galactic clusters are from the spectral fitting
technique by Koleva et al.\ (2008). Metallicities from Schiavon
et al. (2005) for three of these GC are marked by "*".  Col.\ (7) and
(8): Ages and metallicities obtained from CMD studies.  Col.\ (9):
References for Col. 7 and 8. Ke07: Kerber et al.\ (2007); Ge97:
Geisler et al.\ (1997); Ol98: Olsen et al.\ (1998); Gi95: Girardi et
al.\ (1995); Di00: Dirsch et al.\ (2000).  Col.\ (10), and Col.\ (11):
Extinctions derived from CMD studies, Magellanic Clouds Photometric
Surveys (MCPS), respectively, tabulated by Pessev et al.\ (2008).
Col.\ (12): Extinctions values from McLaughlin \& van der Marel
(2005).}
\label{tab:Data}
\end{table*}
%*************************************************************************************************

\section{Evolutionary Synthesis Models}
\label{sec:STPop_Models}

Evolutionary tracks and stellar libraries are the two most important
ingredients in evolutionary synthesis models. Stellar libraries, in
particular, have been the subject of many independent studies over the
past five years. Several groups have put a lot of effort towards
building complete stellar libraries at high/intermediate spectral
resolution, mainly at optical wavelengths (see e.g.\ the review by
Gonz\'alez Delgado 2009). As a result of these efforts, there are
nowadays several sets of models available for spectral synthesis work.

From a neutral user perspective, it is not at all obvious which set of
models should be adopted. Since one of the main goals of this paper is
precisely to evaluate the impact of this choice, we fit each of our SC
spectra with models from four different publicly available sources,
out of which a total of 7128 SSP spectra were retrieved. We are
particularly interested in evaluating the effects of the spectral
libraries, but the chosen models also differ in evolutionary tracks,
IMF and ``model-maker'' (i.e., the evolutionary synthesis code which
combines these ingredients to produce SSP spectra). While
non-exhaustive, this large set is highly representative of the options
currently available to model users.

In what follows we briefly describe each of these sets of models.  A
summary of these models is given in Table 2.

%*******************************************************************************************************************
\begin{table*}
\centering
%  \begin{minipage}{130mm}
\caption{Sets of SSP models used for spectral fitting}
\begin{tabular}{@{}lrccccccrc@{}} \hline
Model-set &  &   Code  &  Isochrones  & IMF  &  Library  & $N_Z$ & Metallicities  &  $N_t$  &  Ages [Gyr]\\  \hline
BC00s   & -1 & Galaxev  & Padova2000 & Salpeter  & STELIB  & 6 & 0.03, 0.019, 0.008, 0.004, 0.001, 0.0004 &  221 & 0--20\\
BC00c   & +1 & Galaxev  & Padova2000 & Chabrier  & STELIB  & 6 & 0.03, 0.019, 0.008, 0.004, 0.001, 0.0004 &  221 & 0--20\\
BC94s   & -2 & Galaxev  & Padova1994 & Salpeter  & STELIB  & 6 & 0.05, 0.02, 0.008, 0.004, 0.0004, 0.0001 &  221 & 0--20\\
BC94c   & +2 & Galaxev  & Padova1994 & Chabrier  & STELIB  & 6 & 0.05, 0.02, 0.008, 0.004, 0.0004, 0.0001 &  221 & 0--20\\
CB94c    & +3 & Galaxev  & Padova1994 & Chabrier  & MILES   & 6 & 0.05, 0.02, 0.008, 0.004, 0.0004, 0.0001 &  221 & 0--20\\
RG00s   & +4 & SED@     & Padova2000 & Salpeter  & GRANADA & 3 & 0.019, 0.008, 0.004                      &   74 & 0.004--17.78\\ 
V00s   & +5 & Vazdekis    & Padova2000 & Salpeter  & MILES   & 6 & 0.03, 0.019, 0.008, 0.004, 0.001, 0.0004 &   46 & 0.1--17.78\\ \hline
\label{tab:STPop_Models_Summary}
\end{tabular}
\end{table*}
%*******************************************************************************************************************

\subsection{Galaxev and STELIB}
\label{sec:STPop_Models_BC03}

STELIB\footnote{http://www.ast.obs-mip.fr/users/leborgne/STELIB/index.html}
is an empirical library that contains 249 stellar spectra in the
3200--9500 \AA\ range with a spectral resolution of 3 \AA\ (Le Borgne
et al.\ 2003). There are 84 stars at solar metallicity, 42 stars are
oversolar and 69 stars around  half solar  and only 23 and 18 stars with metallicity in the
intervals 0.1--0.3, and 0.01--0.1 solar, respectively. The library
lacks hot (over 10000 K), metal-rich, and cool dwarf stars.

Galaxev\footnote{http://www2.iap.fr/users/charlot/bc2003} (BC03)
computes the evolutionary synthesis models with two assumptions for
the IMF: Salpeter and Chabrier (Chabrier 2003), both between 0.1 and
100 $M_\odot$.  Two sets of evolutionary models are used, one with the
Padova 1994 (Bertelli et al.\ 1994) and the other with Padova 2000
(Girardi et al.\ 2000, 2002) isochrones.  The 1994 isochrones are
recommended by BC03, on the basis that the 2000 isochrones produce red
giant branches that are 50--200 K hotter than Padova 1994, and in
consequence they yield older ages for elliptical galaxies.

The SSP spectra are distributed in 221 ages ($0 \le t \le 20$ Gyr) and
6 metallicities ($0.0001 \le Z \le 0.05$ for Padova 1994 and $0.0004
\le Z \le 0.03$ for Padova 2000).  Note that the results obtained with
Galaxev+STELIB models at low and high metallicities are limited by the
low number of stars at $Z \leq 0.004$ and the lack of metal rich giant
stars (see BC03). For SSPs older than 100 Myr and $Z$ between 0.2
solar and solar, these models can produce accurate predictions.

\subsection{SED@ and the GRANADA stellar library}
\label{sec:STPop_Models_GRANADA}

The GRANADA library\footnote{http://www.iaa.csic.es/$\sim$rosa}
(Martins et al.\ 2005) contains 1654 high spectral resolution stellar
spectra with a sampling of 0.3 \AA\ covering from 3000 to 7000
\AA. The spectra were computed for a wide range of effective
temperatures ($3000 \leq T_{\rm eff} \leq 55000$ K), and gravity
($-0.5 \leq \log g \leq 5.5$), and four metallicities (0.1, 0.5, 1 and
2 solar). The most up-to-date stellar atmosphere models were used: a)
the non-LTE line blanketed models from Lanz \& Hubeny (2003) for hot
stars ($T_{\rm eff} \geq 27500$ K); b) ATLAS9 models (Kurucz 1993) for
intermediate temperature stars (4500--27000 K); and c) PHOENIX LTE
line blanketed models (Hauschildt \& Baron 1999) for cool stars
(3000--4500 K). High resolution synthetic spectra were obtained with
the programs SYNTHE and ROT by Hubeny, Lanz \& Jeffery (1995).

The library was implemented in
SED@\footnote{http://www.iaa.csic.es/$\sim$mcs/sed@} (Cervi\~no \&
Luridiana 2006) to predict SSP spectra using the non-rotating Geneva
tracks including standard mass-loss rates and the Padova tracks
(Gonz\'alez Delgado et al.\ 2005). Here, however, we use only the
models generated with the Padova 2000 isochrones (Girardi et al.\
2000, 2002).  The IMF is assumed to be that of Salpeter (1955) between
0.1 and 120 $M_\odot$.  SSP models for 3 metallicities (0.004, 0.008
and 0.019), and 74 ages between 4 Myr and 17.78 Gyr are available.

\subsection{Vazdekis-MILES}
\label{sec:STPop_Models_Vaz}

The MILES library contains about one thousand stars spanning a large
range of stellar parameters (S\'anchez-Bl\'azquez et al.\ 2006). The
spectra were taken at the Isaac Newton Telescope in the Roque de Los
Muchachos Observatory at La Palma, covering the range from 3500 to
7500 \AA\ at a resolution of 2.3 \AA\ (FWHM). This library represents
a significant improvement with respect to STELIB in the coverage of
metallicity, number of giant stars and other aspects, such as flux and
wavelength calibrations. However, MILES has still only a small number
of hot ($> 15000$ K) stars.

This library was incorporated into the evolutionary synthesis code of
Vazdekis (1999). SSP
spectra\footnote{http://www.ucm.es/info/Astrof/miles/models/models.html}
using the Padova 2000 tracks and Salpeter IMF were computed for 6
metallicities in the 0.0004--0.03 range, and 46 ages between 0.1 and 18
Gyr.  The lack of younger ages is due to the lack of hot stars in
MILES.

\subsection{Galaxev and MILES/GRANADA}
\label{sec:STPop_Models_CB07}

Charlot \& Bruzual (2009 in prep.) have produced models analogous to
those in BC03, but replacing STELIB by a combination of the MILES and
GRANADA libraries. The latter are used only for hot stars.  These
models, kindly provided by the authors in advance of publication, are
still preliminary. The models used here have a Chabrier IMF and Padova
1994 tracks, with the same 221 ages and 6 metallicities in BC03.

\subsection{Summary of the set of models}
\label{sec:STPop_Models_Summary}

Table \ref{tab:STPop_Models_Summary} summarizes the basic properties
of the different sets of SSP models used in this work.  There are 7
sets of models, differing in at least one of the following:
evolutionary synthesis code (Galaxev, SED@, and Vazdekis), stellar
tracks (Padova 1994 and 2000), IMF (Chabrier, Salpeter) and stellar
library (STELIB, MILES, GRANADA). Independent combinations of these
four ingredients would produce a grid of $3 \times 2 \times 3 \times
3 = 54$ model-sets, but this is not how one finds these models in
their respective web repositories. Furthermore, the sets of
metallicities are not identical among all models, and there are also
differences in the ages for which SSP spectra are tabulated.
Comparisons of results achieved with different models must take into
account this non-uniformity, which is an example of one of the
difficulties faced by model users.

The spectral fitting methodology described in Paper I was applied to
all 27 SC spectra. As explained there, the fits are carried out
feeding {\sc starlight} with a single $Z$ base containing all $N_t$
ages in a model, and then repeating things for other $Z$'s. The $N_Z =
6$ V00s bases in Paper I expand to 39 bases to account for all models
in Table \ref{tab:STPop_Models_Summary}.  Note that while the
sampling in age is practically continuous, predictions are only
available for 6 metallicities (3 in the case of the GRANADA models),
an apparent ``technicality'' which has a non-negligible impact in our
analysis, and is not fully cured by interpolations (see Paper I).

\subsection{Notation}

Throughout this paper these models will be referred to using the codes
given in Table 2.  Alternatively, we will refer to BC as the STELIB
models, CB and V as the MILES models, and RG as the GRANADA
models. For convenience, metallicities will be transformed to a
log-solar scale.  In this notation, the $Z$ values 0.05, 0.03, 0.019,
0.008, 0.004, 0.001, 0.0004 and 0.0001 covered by the models
correspond to (rounding up to the first decimal) $\log Z/Z_\odot =
+0.4$, +0.2, 0, -0.4, -0.7, -1.3, -1.7 and -2.3, respectively. (Notice
that we do not distinguish between $Z = 0.019$ and 0.020).  This will
be applied both to the metallicities from the stellar tracks and
[Fe/H] from the literature. Ages will be given in yr throughout.

\section{Results: Quality of the spectral fits}
\label{ref:SpectralFits}

%***FIG***FIG***FIG***FIG***FIG***FIG***FIG***FIG***FIG***FIG***
\begin{figure*}
\includegraphics[width=\textwidth]{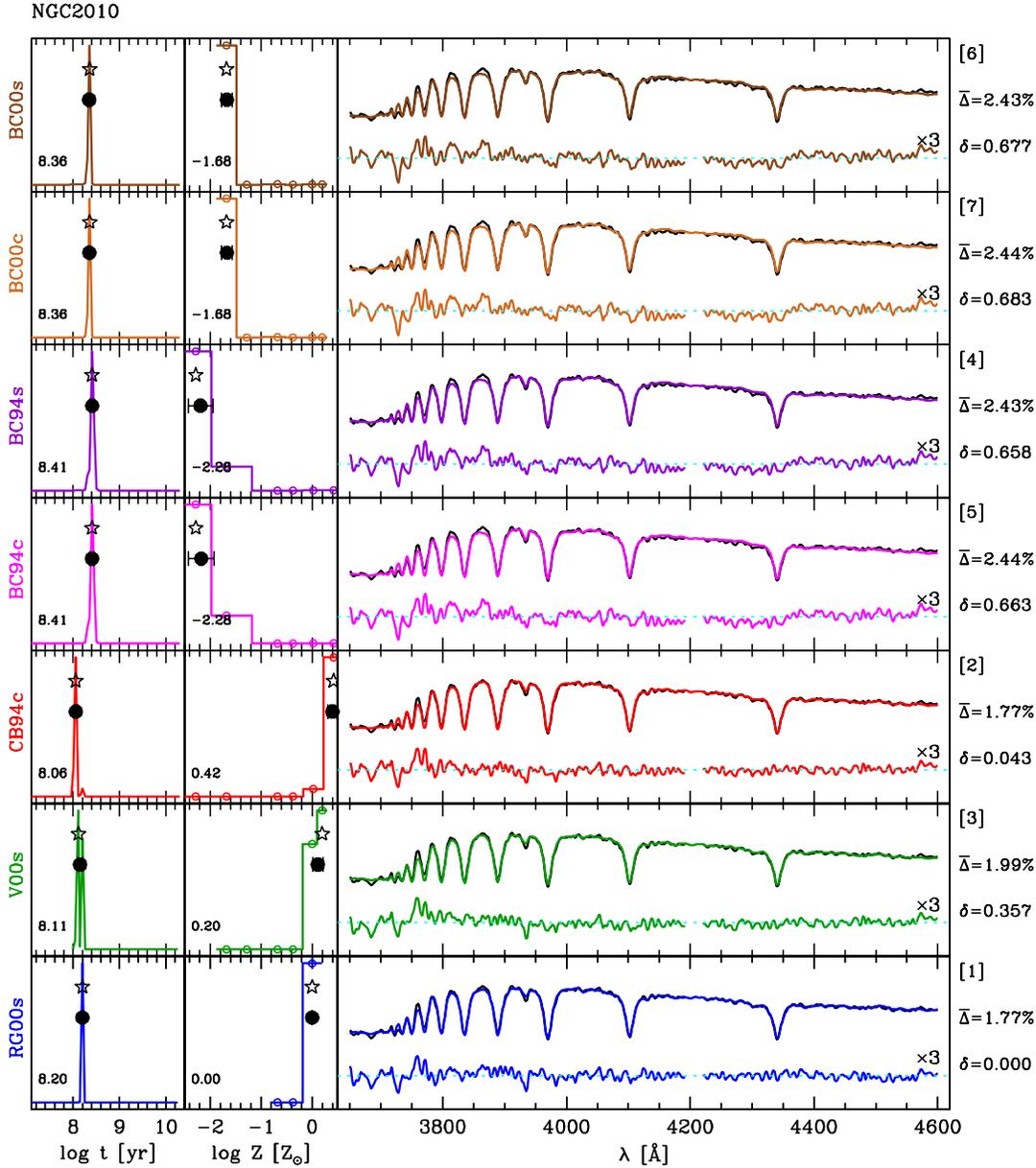}
\caption{Comparison of spectral fits of NGC 2010 obtained with 7
different evolutionary synthesis, labeled in the left axis (see Table
\ref{tab:STPop_Models_Summary}). Right panels show the observed
($O_\lambda$, in black), best fit ($M_\lambda$, colored), and
$R_\lambda = O_\lambda - M_\lambda$ residual spectra. All spectra are
normalized at 4020 \AA, and the dashed line marks the zero flux level.
{\em Notice that $R_\lambda$ is multiplied by 3 for clarity}. Values
of the mean percentage residual ($\overline{\Delta}$) are listed to
the right of each panel, where the number in brackets indicate the
$\chi^2$ fit quality ranking, and $\delta = (\chi^2 - \chi^2_{best}) /
\chi^2_{best}$. The left panels show the probability distribution
functions of $t$ and $Z$.  A solid circle with error bars marks the
mean $\pm 1$ sigma estimates. The numbers correspond to the best fit
$t$ (left most panel) and $Z$ (middle), whose values are also marked
by a star.  In the middle panel, open circles are plotted in each of
the metallicities in the base.}
\label{fig:AllFits_NGC2010}
\end{figure*}

\begin{figure*}
\includegraphics[width=\textwidth]{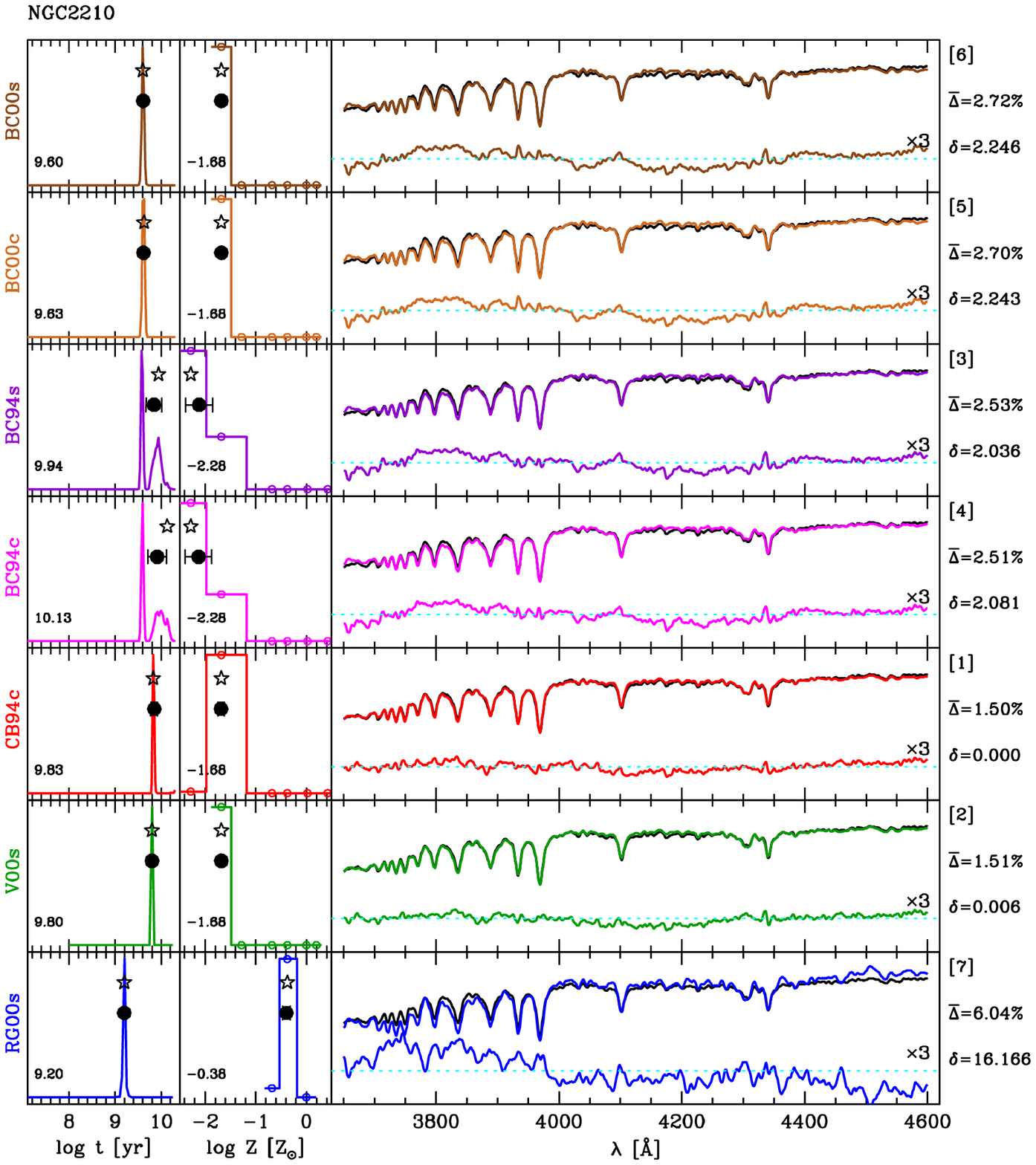}
\caption{As Fig.\ \ref{fig:AllFits_NGC2010}, but for NGC 2210}
\label{fig:AllFits_NGC2210}
\end{figure*}
%***FIG***FIG***FIG***FIG***FIG***FIG***FIG***FIG***FIG***FIG***

As a first step in the exploration of our results, this section
investigates the quality of the spectral fits obtained with different
models. Given the size of the data set (27 SCs $\times$ 7 models), we
concentrate this comparison on results for a couple of illustrative
examples (the same used in Paper I), which also help clarifying
several of the issues discussed in the next sections.

Fig.\ \ref{fig:AllFits_NGC2010} shows the spectral fits to the LMC
cluster NGC 2010. To first order all fits (right panels) are of
similar quality, as can be appreciated by the similarity of the
residual spectra plotted (multiplied by 3 for clarity). The mean
percentage deviations ($\overline{\Delta}$) vary from 1.8 (RG00s) to
2.4\% (BC00s), a relatively narrow range which shows that, at least in
the case of this young metal rich SC, it is hard to distinguish models
on the basis of fit quality alone.

As discussed in Paper I, our $\chi^2$ values are not in a meaningful
absolute scale, but can be used to rank the fits. In analogy with the
index defined in Paper I to compare fits of different $t$, $Z$, and
$A_V$ but same model set, we will use

\begin{equation}
\label{eq:delta}
\delta = \frac{\chi^2 - \chi^2_{best}}{\chi^2_{best}}
\end{equation}

\noindent to compare fits obtained with different models, with $\delta
= 0$ denoting the best one (RG00s in the case of NGC 2010).

Values of $\delta$ are listed in Fig.\ \ref{fig:AllFits_NGC2010},
along with the resulting model ranking (numbers within brackets). The
$\delta$ values for NGC 2010 show that RG00s and CB94c models are
nearly equally good, with the V00s models coming in third place. The
STELIB based models not only produce the worse spectral fits, but
their $t$ and $Z$ values differ a lot from the others, as can be seen
in the probability distribution functions (PDF) plotted on the left
and middle panels.

Moving to the other end of the $t$-$Z$ space, results for the old and
metal poor cluster NGC 2210 are shown in Fig.\
\ref{fig:AllFits_NGC2210}. In this case the CB94c and V00s models are
visibly much better than the others. This is reflected by the $\delta$
values, which are $> 2$ for all other models (indicating fits over 3
times as worse in terms of $\chi^2$).  The RG00s models, which
performed so well for NGC 2010, are by far the worst in this case.
The reason for this is that the RG00s models start at $\log Z/Z_\odot
= -0.7$ which is much larger that the metallicity of this SC (Table
\ref{tab:Data}). This also happens for NGC 416, NGC 1754, M15 and M79.
Besides, it is well known that theoretical libraries perform
better for hot than for cool stars (e.g. Martins \& Coelho 2007;
Bertone et al.\ 2008), so one expects the GRANADA models to work
better for young than for old SCs, irrespective of their metallicity.
This is confirmed by the case of 47 Tuc. With $\log Z/Z_\odot = -0.76$
(Table \ref{tab:Data}), this SC is not too far off the lower $Z$ limit
of the GRANADA models, but because of its old age ($\log t = 10.1$),
the RG00s models provide the worse spectral fits (Tables
\ref{tab:Results_BCBMmc}--\ref{tab:Results_BBC00s}).

These examples suggest that, at least in some cases, spectral fitting
can, by itself, help distinguish among competing models.  Another
conclusion of this preliminary analysis is that models based on the
MILES library tend to produce better spectral fits than those based on
STELIB. The same can be said about the GRANADA models if one
concentrates on young and intermediate age SCs.  This is rewarding in
the sense that it shows that more recent libraries (MILES and GRANADA)
do represent an improvement upon older ones (STELIB), at least insofar
as spectral fit quality is concerned and as long as the limitations of
the models are not overlooked.

\begin{figure}
\includegraphics[width=0.5\textwidth]{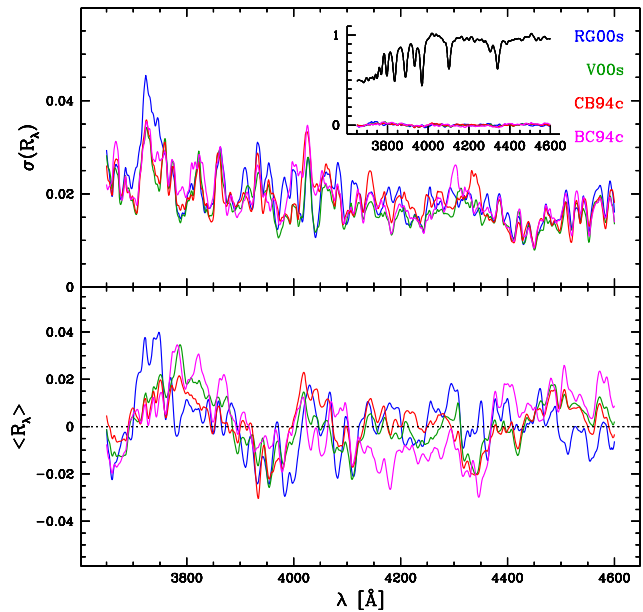}
\caption{Mean (bottom) and standard deviation (top) of the spectral
residuals for the 13 SCs in the sample whose ages and metallicities
are within the ranges spanned by all models: NGC 1651, NGC 1783, NGC
1806, NGC 1831, NGC 1846, NGC 1978, NGC 2010, NGC 2133, NGC 2134, NGC
2136, NGC 2203, NGC 2213 and NGC 2249. The inset shows the same
residuals as in the bottom panel, but plotted on the scale of the mean
observed spectrum (black line). All spectra have been smoothed by 10
\AA\ boxcar for clarity.}
\label{fig:SpecResidStats}
\end{figure}

On the whole, however, these improvements are relatively subtle, 
only marginally captured by global fit quality measures. For instance,
restricting to 13 SCs whose $t$ and $Z$ literature values fall within
the nominal range of validity of all models considered here, we find
sample mean values of $\overline{\Delta} = 2.15$, 2.19 and 2.38\% for
V00s, CB94c and RG00s models, respectively, while for the STELIB
models the values range from 2.42 to 2.51\%.  These small differences
are further illustrated in Fig.\ \ref{fig:SpecResidStats}, where we
plot the mean residual spectrum (bottom) and its standard deviation
(top) for the same subset of 13 SCs.  For clarity, all spectra have
been smoothed by 10 \AA. Because of its vertical scale, the plot may
convey the wrong impression that residuals are large. To show that
this is not the case, the inset shows the same mean residuals as in
the bottom panel, but plotted on the more relevant scale of the mean
spectrum.

The general message from Fig.\ \ref{fig:SpecResidStats} and the sample
mean $\overline{\Delta}$ values quote above is the same.  They both
confirm that progress is being made, but they also show that, from the
point of view of fit quality, these are small improvements upon models
which were already good. This highlights the need to extend these
tests from the space of observables (spectra) to that of derived
properties ($t$, $Z$ and $A_V$). That is the subject of the next
sections.

%***FIG***FIG***FIG***FIG***FIG***FIG***FIG***FIG***FIG***FIG***
%***FIG***FIG***FIG***FIG***FIG***FIG***FIG***FIG***FIG***FIG***

\section{Results: Comparison of properties derived with different models}
\label{sec:Results_ModXMod}

We now turn to a comparison of the SC properties derived from spectral
fits with different models. Our goal here is to evaluate the
dispersion and systematic differences in $t$, $Z$ and $A_V$ stemming
from the use of different ingredients in the spectral analysis.

For each of $t$, $Z$, and $A_V$, the methodology outlined in Paper I
provides two estimates: The best fit value and the mean value obtained
from the full PDF; subscripts 'best' and 'PDF' are used to distinguish
them when necessary.  Both estimates are shown in the PDF panels in
Figs.\ \ref{fig:AllFits_NGC2010} and \ref{fig:AllFits_NGC2210}, where
stars mark $t_{\rm best}$ and $Z_{\rm best}$ and solid circles with an
error bar show the PDF based estimates.

Tables \ref{tab:Results_BCBMmc}--\ref{tab:Results_BBC00s} give these
estimates for all SCs and 6 of the 7 models considered here except for
V00s, whose results were already presented in Table 1 of Paper I.
Other entries in these tables give the mean percentage deviation
($\overline{\Delta}$) of the best fit model and quantities related to
the multi-SSP fits (see Paper I for details).  From the discussion in
Paper I (see also Koleva et al.\ 2008), in a few cases (NGC 2210, M15,
M79) multi-SSP fits detect the presence of a hot and old population
not well accounted for by single SSP models, indicating the presence
of a blue Horizontal Branch (HB). In such cases, the age of the oldest
component (column 10 in Tables
\ref{tab:Results_BCBMmc}--\ref{tab:Results_BBC00s}) is arguably a
better age estimate than either $t_{\rm best}$ or $t_{\rm PDF}$. In
the case of NGC 2210 and for the V00s models, for instance, this
third estimate gives an age $\sim 0.3$ dex older than that found with
single-SSP fits, and in better agreement with the literature data.

We concentrate our analysis on the PDF-based (or ``bayesian'')
estimates of $t$, $Z$ and $A_V$. Barring caveats like systematic
effects associated to SCs with blue HB's, we consider these our more
consistent estimates of SC properties.

\begin{figure}
\includegraphics[width=0.5\textwidth]{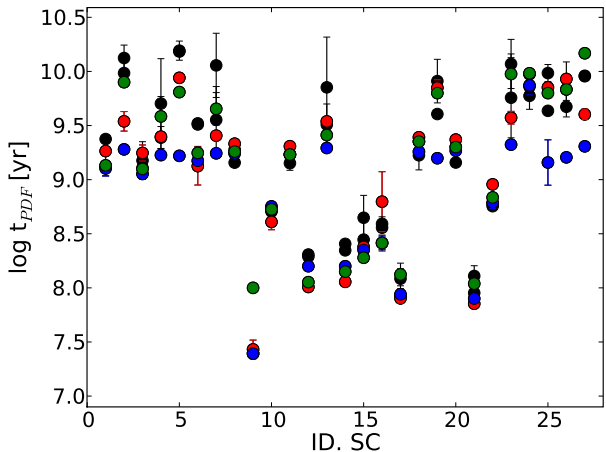}
\caption{Age estimates obtained from the spectral fits of each cluster
for each set of evolutionary synthesis models, coded by color: STELIB
models (only BC00s and BC94c) results are plotted in black, CB94c
models in red, RG00s models in blue, and V00s models in green
circles. Every point has an error bar, but only the STELIB (black)
ones stand out clearly.}
\label{fig:ModXMod_age}
\end{figure}

\begin{figure}
\includegraphics[width=0.5\textwidth]{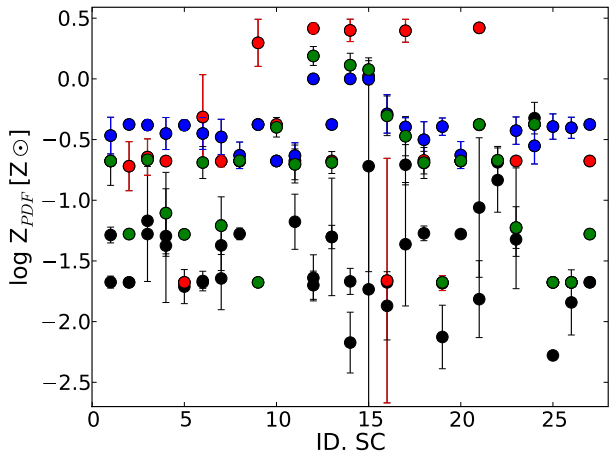}
\caption{As Fig.\ \ref{fig:ModXMod_age}, but for the metallicity.}
\label{fig:ModXMod_Z}
\end{figure}

\begin{figure}
\includegraphics[width=0.5\textwidth]{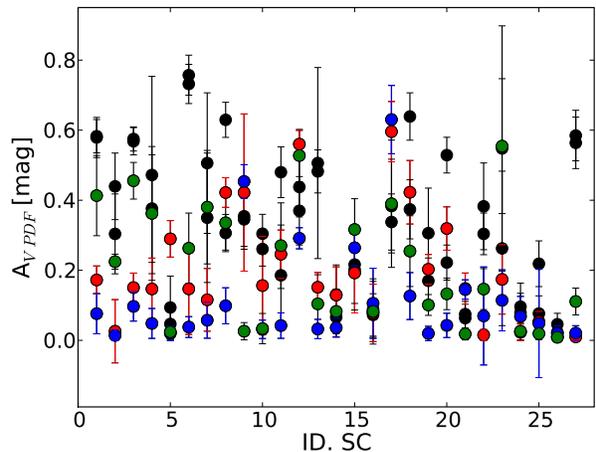}
\caption{As Fig.\ \ref{fig:ModXMod_age}, but for the extinction.}
\label{fig:ModXMod_AV}
\end{figure}

Figs.\ \ref{fig:ModXMod_age}, \ref{fig:ModXMod_Z} and
\ref{fig:ModXMod_AV} plot $t_{\rm PDF}$, $Z_{\rm PDF}$ and $A_{V, \rm
PDF}$ for all models and SCs in the sample.  Each SC is identified by
an ID number (see Table \ref{tab:Data}), and each set of models is
coded by a color.  One should note that formal parameter
uncertainties (the PDF-based standard deviations, plotted as error
bars) are themselves subjected to uncertainties and biases due to
model limitations. The fact that the GRANADA models (blue points) do
not extend to metallicities as small as that of many of our SCs leads
to underestimated $\sigma(\log Z/Z_\odot)$ values. Similarly, because
of the lack of $t < 10^8$ yr models in the Vazdekis grid, the derived
$\sigma(\log t)$ values are too small for the youngest SCs.  In any
case, even taking these caveats into consideration, these plots show
that formal statistical uncertainties are smaller than the differences
stemming from use of different models.  In what follows we quantify
and discuss the reasons for these differences.

%***FIG***FIG***FIG***FIG***FIG***FIG***FIG***FIG***FIG***FIG***
%***FIG***FIG***FIG***FIG***FIG***FIG***FIG***FIG***FIG***FIG***

%***FIG***FIG***FIG***FIG***FIG***FIG***FIG***FIG***FIG***FIG***
%***FIG***FIG***FIG***FIG***FIG***FIG***FIG***FIG***FIG***FIG***

%***FIG***FIG***FIG***FIG***FIG***FIG***FIG***FIG***FIG***FIG***
%***FIG***FIG***FIG***FIG***FIG***FIG***FIG***FIG***FIG***FIG***

\subsection{Ages} 

Fig.\ \ref{fig:ModXMod_age} shows, for each SC, the bayesian age
obtained with each of the seven models.  For most objects, the formal
uncertainty in $\log t$ is small, indicating that the fit is well
constrained around a given combination of $t$ and $Z$ in the base
models.

%***FIG***FIG***FIG***FIG***FIG***FIG***FIG***FIG***FIG***FIG***
\begin{figure}
\includegraphics[width=0.5\textwidth]{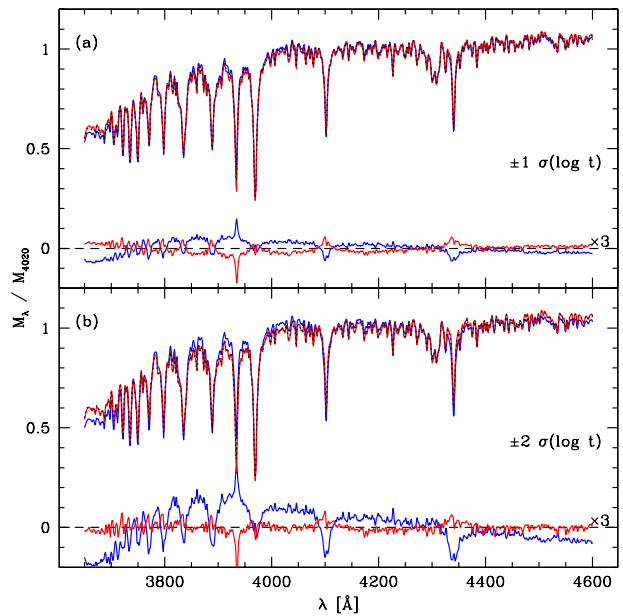}
\caption{Comparison of CB94c models of different ages for
NGC2210. Dotted lines show the model whose age is the closest to our
PDF-based estimate ($\log t = 9.86$). Blue and red lines in panel (a)
show the models at ages one $\sigma(\log t)$ above and below $\log
t_{\rm PDF}$, while in (b) models with ages $\pm 2 \sigma(\log t)$
from $\log t_{\rm PDF}$ are plotted. Models are shown at their
original spectral resolution, and residuals are multiplied by 3 for
clarity. The plot illustrates the sort of spectral differences
expected for parameters within 1 and 2 sigma of their estimated
values.}
\label{fig:JesusFig}
\end{figure}
%***FIG***FIG***FIG***FIG***FIG***FIG***FIG***FIG***FIG***FIG***

To provide a visual sense of the significance of our PDF-based
uncertainties in age, Fig.\ \ref{fig:JesusFig} shows CB94c models for
NGC 2210 for 5 ages: the PDF estimate $\log t = 9.86$ (dotted lines),
plus models with ages one (top) and two (bottom) sigmas away from this
value. Since $\sigma(\log t) = 0.09$ (Table \ref{tab:Results_BCBMmc}),
the models shown span the $9.68 < \log t < 10.04$ interval. All SSP
spectra correspond to $\log Z/Z_\odot = -1.7$, but $A_V$ varies from 0.26 (at
$\log t_{\rm PDF} - 2 \sigma$) to 0.12 (at $\log t_{\rm PDF} + 2
\sigma$), in a way to always produce the best possible match to the
observed spectrum.  The plot shows that models are indeed
spectroscopically different even at a $\pm 1 \sigma(\log t)$ level,
which gives qualitative support to our formal error estimates.  It is
nevertheless necessary to point out that in many cases the quality of
the fit deteriorates so quickly even for adjacent ages in the grid
that $\sigma(\log t)$ becomes $< 0.01$, in which case no uncertainty
is quoted (these are the $\pm 0.00$ entries in Tables
\ref{tab:Results_BCBMmc}--\ref{tab:Results_BBC00s}). This happens
often in the V00s models, as a result of their relatively coarse (0.05
dex) sampling in age. As pointed out in Paper I, in such cases it is
more advisable to take half of the grid sampling as a more realistic
(albeit less formal) estimate of $\sigma(\log t)$.

\begin{figure}
\includegraphics[width=0.5\textwidth]{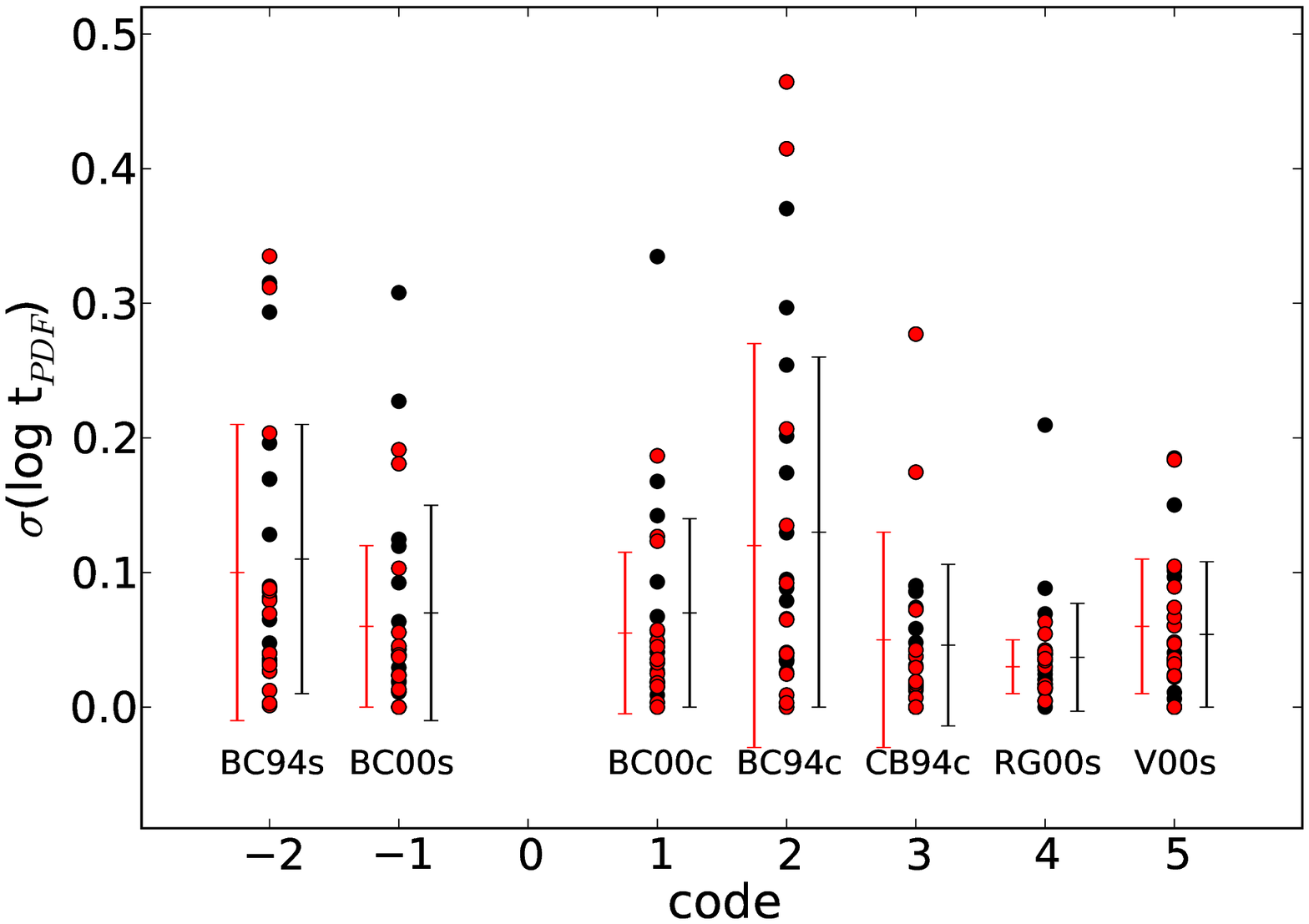}
\includegraphics[width=0.5\textwidth]{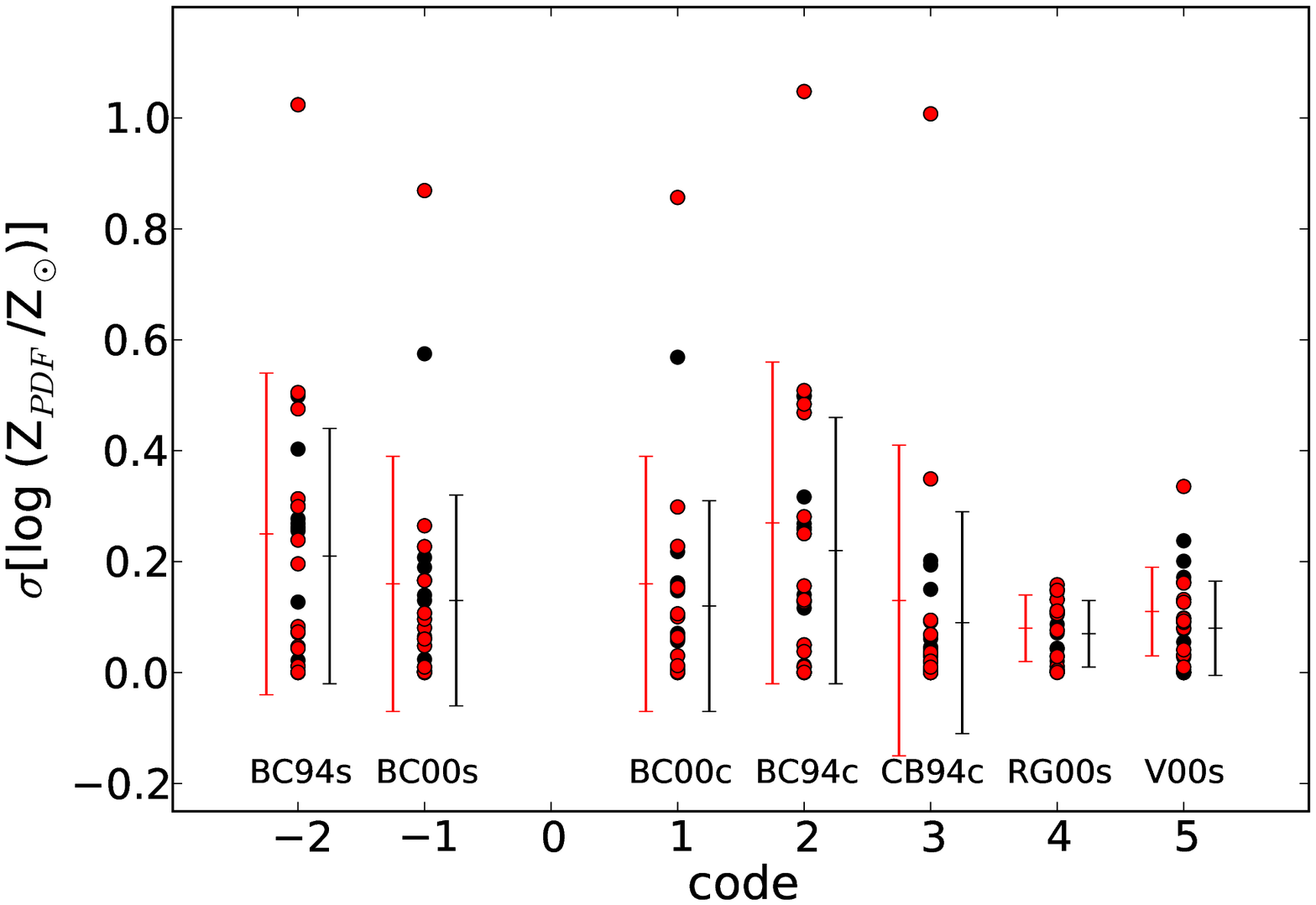}
\caption{Distributions of $\sigma(\log t)$ and $\sigma(\log
Z/Z_\odot)$ versus the set of models used in the fits.  Each point
represents the result obtained for each cluster and each set of
models. The error bars represents the mean and the standard deviation
of the distribution for each set of models.  Points and error
bars plotted in red correspond to the subset of 13 clusters which fall
within the range of validity of all models considered here (the same
ones listed in Fig.\ \ref{fig:SpecResidStats}).  }
\label{fig:Age_and_Z_sigma}
\end{figure}

In some cases, however, $\sigma(\log t)$ is relatively large.  The
broadest age PDFs (larger uncertainty in $t$) are found with STELIB
models (see Fig.\ \ref{fig:Age_and_Z_sigma}). The example of NGC 2210,
whose $t$ and $Z$ PDFs are shown in Fig.\ \ref{fig:AllFits_NGC2210},
illustrates that such large values for $\sigma(\log t)$ stem from the
inability to distinguish among different $Z$'s. The PDF($Z$) for the
BC94 fits is broad, covering the 2 smallest values in the grid ($\log
Z/Z_\odot = -2.3$ and $-1.7$). To each of these two acceptable $Z$'s
there is a corresponding peak in PDF($t$), producing a large
$\sigma(\log t)$\footnote{The BC00 models would most likely exhibit
the same behavior if their $Z$'s reached values as low as in BC94 (the
lowest $\log Z/Z_\odot$ in the BC00 models is $-1.7$).}.  This
ambiguity arises from the fact that STELIB contains very few truly low
$Z$ stars, so the predicted SSP spectra for these two $Z$'s use
essentially the same stars, differing only in their proportions
(dictated by the evolutionary tracks). This leads to similar predicted
SSP spectra, explaining the inability of spectral fits to distinguish
the two solutions. In other words, there is a severe mismatch between
the metallicities of the stars in the library and the nominal $Z$ of
the models, which is that corresponding to the evolutionary
tracks. This caveat is acknowledged and discussed by BC03, who
classify their two lowest metallicity models as ``poor'' for $< 1$ Gyr
populations and ``fair'' for older ones (compare this to the ``very
good'' mark ascribed to $Z_\odot$ models of any age).

A similar conclusion arises when $t_{\rm PDF}$ is plotted against
$t_{\rm best}$. Naturally, these two quantities correlate strongly,
but the few points out of the correlation correspond to STELIB fits,
and these outliers are also points that have the largest $\sigma(\log
t)$.  Thus, ages are more poorly constrained with the STELIB models.

\begin{figure*}
  \centering
%  \begin{minipage}{140mm}
\includegraphics[width=\textwidth]{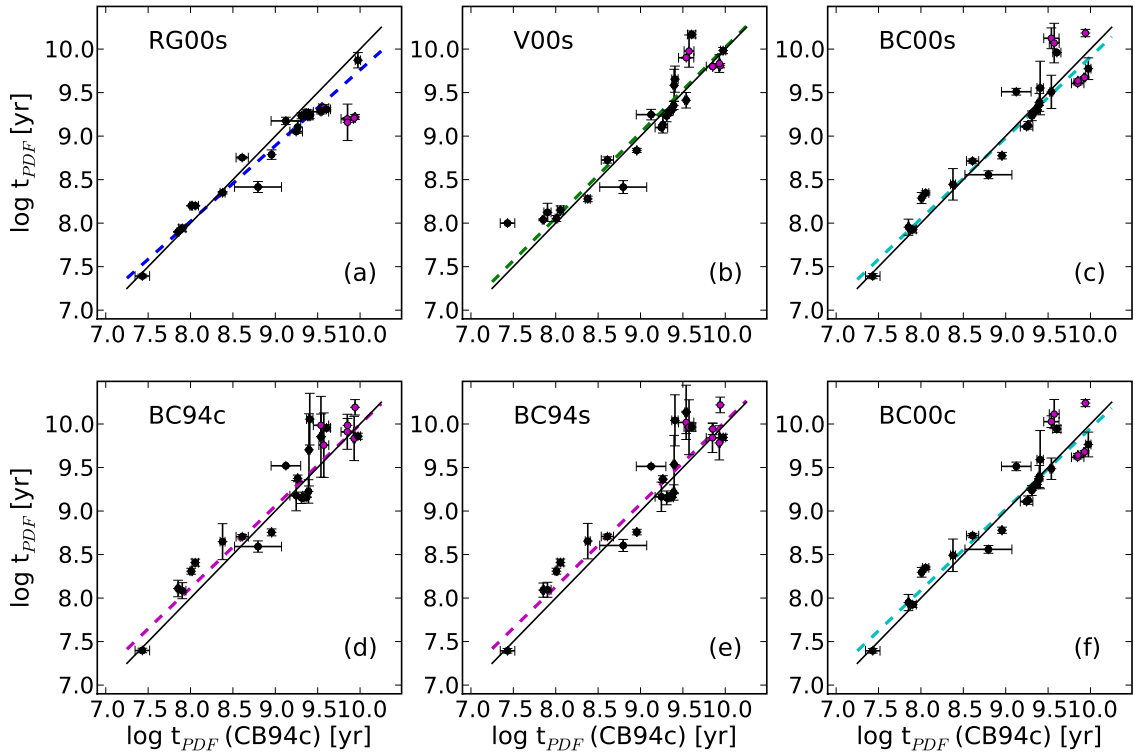}
\caption{Correlation between the bayesian ages obtained for each
set of models as a function of the ages obtained with CB94c models.
The black line is the one-to-one relation, and the dashed line shows
a linear fit.}
\label{fig:ageagemodels}
% \end{minipage}
\end{figure*}

Notwithstanding these caveats, the age estimates obtained with all
models are fairly consistent with each other. This is illustrated in
Fig.\ \ref{fig:ageagemodels}, where we plot the bayesian ages of
RG00s, V00s and BC models against those obtained with the CB94c
models.

The largest discrepancies seen in Figs.\ \ref{fig:ModXMod_age} and
\ref{fig:ageagemodels} are associated to fits of SCs whose ages or
metallicities fall outside the range of validity of the models,
causing expected biases.  NGC 1818 (ID.\ 9 in Fig.\
\ref{fig:ModXMod_age}), for instance, has a CMD age of 20 Myr (Table
\ref{tab:Data}).  All the models, except V00s, give an age in
agreement with the CMD age. The reason why the V00s models fail for
this SC is simply that their youngest age is 100 Myr, which is in fact
the best fitting age found by {\sc starlight} (for lack of a better
alternative).  NGC 1754 (ID. 5 in Fig.\ \ref{fig:ModXMod_age}) is
another example.  Most of the models obtain an age which is similar to
the CMD value ($\log t = 10.2$), but the GRANADA models yield an age
of $\log t = 9.2$, a factor of 10 too low.  This failure occurs
because the lowest metallicity available in the RG00s models is $\log
Z/Z_\odot = -0.7$, much higher than the metallicity of NGC 1754 ($\log
Z/Z_\odot = -1.4$, according to Table \ref{tab:Data}).  Due to the
age-metallicity degeneracy, forcing a larger $Z$ results in an
underestimated $t$. A similar effect is seen in Fig.\
\ref{fig:AllFits_NGC2210}, where the RG00s models produce the smallest
ages.

In short, the RG00s models are simply not applicable to metal poor
SCs, and the V00s cannot be applied to systems younger than 100 Myr.
Except for such violations of the range of validity, all models
provide reasonably consistent age estimates.

Because our sample is composed of well studied objects, with plenty of
$t$ and $Z$ estimates in the literature, it is straightforward to
identify such violations. Obviously, such information is not generally
available for other systems, and in fact this is precisely what one
aims to derive through spectral synthesis.  It is therefore useful to
show results obtained stretching models beyond their nominal range of
validity, as done above. This ``deliberate mistake'' produces
illustrative examples of the sort of spurious results that one would
obtain in practical work, where the same mistake can be made
unadvertedly.

%***FIG***FIG***FIG***FIG***FIG***FIG***FIG***FIG***FIG***FIG***
%***FIG***FIG***FIG***FIG***FIG***FIG***FIG***FIG***FIG***FIG***

%***FIG***FIG***FIG***FIG***FIG***FIG***FIG***FIG***FIG***FIG***
%***FIG***FIG***FIG***FIG***FIG***FIG***FIG***FIG***FIG***FIG***

\subsection{Metallicity}

Fig.\ \ref{fig:ModXMod_Z} shows the metallicities obtained for each
cluster with the seven sets of models. In contrast with the age
results, the dispersion in the metallicity is significant, in most
cases much larger than the error bars.

\begin{figure}
\includegraphics[width=0.5\textwidth]{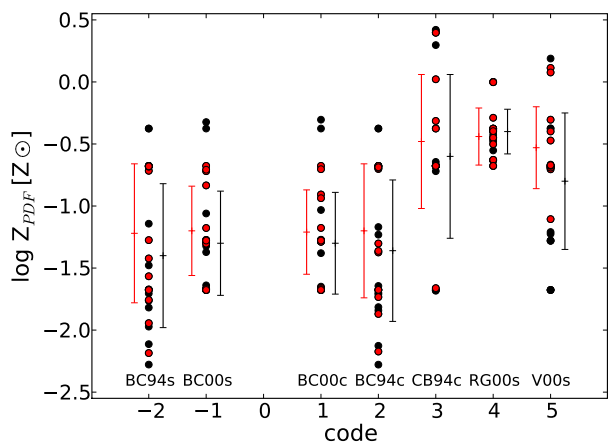}
\caption{Distribution of the metallicity ($\log Z/Z_\odot$) versus the
set of models (code) used. The error bars represent the mean and the
standard deviation of the log (Z/Z$_\odot$) distribution for each set
of models considering the whole sample of clusters (black) and
only 13 clusters in Figure 3 (red).}
\label{fig:Metallicity}
\end{figure}

One source for this large dispersion is the already mentioned
violation of the limits of the RG00s and V00s models.  As expected,
metal poor SCs (IDs 2, 5, 12, 21, 25, and 26 in \ref{fig:ModXMod_Z},
which have $\log Z/Z_\odot \leq -1.2$) are badly fitted with the
GRANADA models, which are valid only for $\log Z/Z_\odot \ge
-0.7$. This also explains why the $Z$ values for the RG00s fits cover
such a narrow range (see also Fig.\ \ref{fig:Metallicity}). Also, the
V00s value of the metallicity of NGC 1818 is severely underestimated
to compensate for its overestimated age of 100 Myr, the youngest SSP
in these models.

Even disconsidering these cases, the dispersion in $Z$ values remains
large. As evident to the eye in Fig.\ \ref{fig:ModXMod_Z}, much of
this dispersion is caused by the STELIB results, which in most cases
produce $Z$ values well below those obtained with other models. Most
of the STELIB results are below 0.1 solar, and none are solar or
above.  As seen in Fig.\ \ref{fig:Metallicity}, the STELIB metallicity
distribution peaks at $\log Z/Z_\odot = -1.6$, while other models peak
around $-0.7$.  Besides this bias, STELIB models also lead to larger
formal uncertainties in $Z$, as shown in Fig.\
\ref{fig:Age_and_Z_sigma}. The reason for this is the already
discussed similarity of SSP spectra for the lowest $Z$'s in BC03.
These results remain valid restricting the analysis to 13 SCs
whose t and Z literature values fall within the nominal range of
validity of all models considered here (the red points in Fig.\
\ref{fig:Metallicity} correspond to the SCs in Fig.\
\ref{fig:SpecResidStats}).

It is important to point out that although throughout this paper we
group the BC94 and BC00 models as ``STELIB models'', there is an
important, albeit apparently technical, difference between them: The
BC94 models have a coarser but wider $Z$-grid, extending down to $\log
Z/Z_\odot = -1.7$ and $-2.3$, whereas the BC00 models stop at $\
-1.7$.  The similarity between the $\log Z/Z_\odot = -1.7$ and $-2.3$
spectra in the BC94 models leads to larger $t$ and $Z$ uncertainties
than obtained with the BC00 models, as can be seen in Figs.\
\ref{fig:Age_and_Z_sigma} and \ref{fig:Metallicity}. Had we ignored
the $\log Z/Z_\odot = -2.3$ BC94 models, these differences in
$\sigma(\log t)$ and $\sigma(\log Z/Z_\odot)$ would be smaller, but
still larger than those obtained with GRANADA and MILES models.

%***FIG***FIG***FIG***FIG***FIG***FIG***FIG***FIG***FIG***FIG***
%***FIG***FIG***FIG***FIG***FIG***FIG***FIG***FIG***FIG***FIG***

\subsection{Extinction}

\begin{figure}
\includegraphics[width=0.5\textwidth]{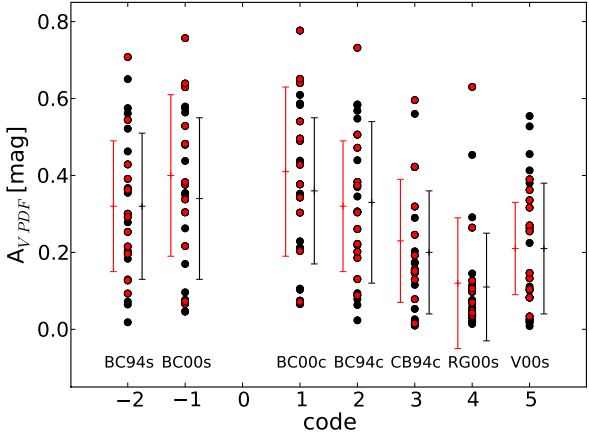}
\caption{As Fig.\ \ref{fig:Metallicity}, but for the V-band
extinction, $A_V$.}
\label{fig:Extinction}
\end{figure}

Fig. \ref{fig:ModXMod_AV} shows for each SC the bayesian extinction
obtained with each set of models.  The values obtained range from 0 to
0.8, with significant model to model differences.  As it happens with
$Z$, the STELIB results for $A_V$ differ systematically from those
derived with the MILES and GRANADA models, as better seen in Fig.\
\ref{fig:Extinction}.  The average extinction with STELIB, MILES and
GRANADA libraries is 0.34, 0.20, and 0.11, respectively.  The
uncertainty in $A_V$ is also larger with STELIB, which produces an
average $\sigma(A_V)$ of 0.1, whereas for other models the average
$\sigma(A_V)$ is 0.06.

%***FIG***FIG***FIG***FIG***FIG***FIG***FIG***FIG***FIG***FIG***
%***FIG***FIG***FIG***FIG***FIG***FIG***FIG***FIG***FIG***FIG***

\section{Results: Spectral fitting results vs.\ data in the literature}
\label{sec:Synthesis_X_Literature}

Despite their relevance, the internal comparisons performed in
section \ref{ref:SpectralFits}  cannot,
by definition, provide an absolute measure of the adequacy of the
results achieved with different models. This section presents this
most critical test. We compare the ages, metallicities and extinctions
derived from spectral fits with those reported in the literature. We
also discuss the capabilities and limitations of each of seven sets of
evolutionary synthesis models to reproduce the results from S-CMD work
or with Rose's indices.

\subsection{Ages}

%***FIG***FIG***FIG***FIG***FIG***FIG***FIG***FIG***FIG***FIG***
\begin{figure*}
\centering
\begin{minipage}{160mm}
\includegraphics[width=\textwidth]{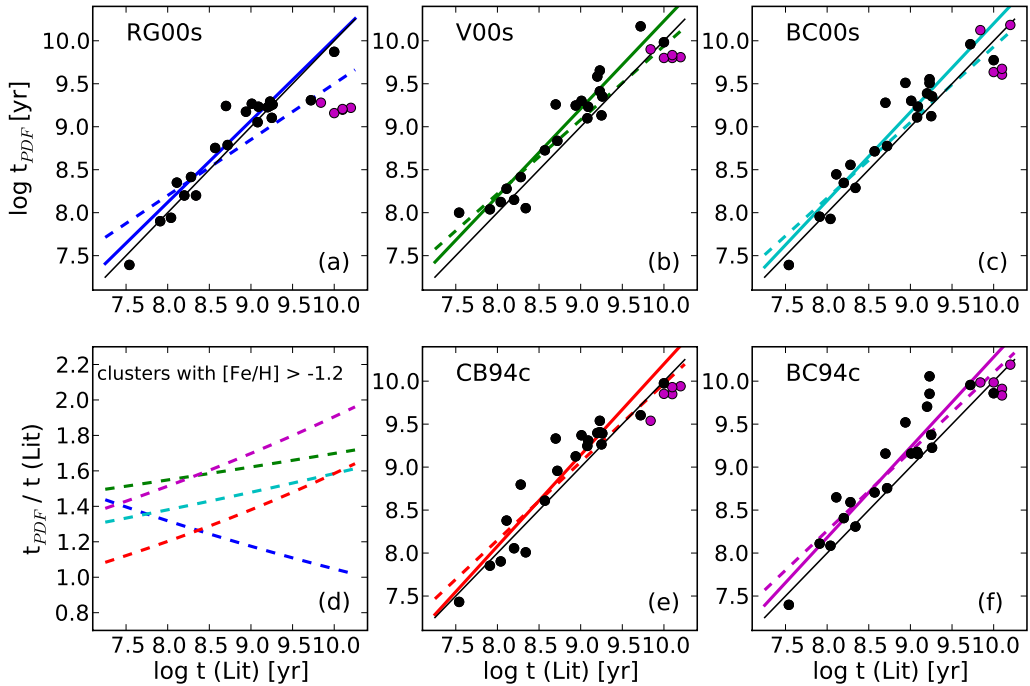}
\caption{Panels a, b,c, e and f: Correlation between the age obtained
with {\sc starlight} and the ages in Table \ref{tab:Data} derived with
the S-CMD calibration.  Each panel represents the result for each set
of models as labelled in the upper left corner.  Metal poor clusters
($\log Z/Z_\odot \leq -1.2$) are marked with magenta circles. The
black line is the one-to-one relationship, the dashed line is the
result of the linear fit to all the points, and the solid line the
result of the linear fit excluding the metal poor clusters.  The
bottom left panel shows the deviation of the linear fit obtained for
each set of models with respect to the one-to-one relationship,
expressed as the ratio between the spectral fit and literature ages.
}
\label{fig:Ages_correlations}
\end{minipage}
\end{figure*}
%***FIG***FIG***FIG***FIG***FIG***FIG***FIG***FIG***FIG***FIG***

Fig.\ \ref{fig:Ages_correlations} compares our PDF based ages with the
S-CMD ages compiled by LR03, listed in Table \ref{tab:Data}.  Each
panel shows results obtained for one of the seven sets of models in
Table \ref{tab:STPop_Models_Summary}. In all cases the correlation is
very good. Two linear fits are presented in each panel: Dashed lines
show the fits obtained using all 27 SCs, while the solid lines
represent fits excluding the five SCs with $\log Z/Z_\odot \le -1.2$
(NGC 416, NGC 1754, NGC 2210, M15 and M79).  The latter fits filter
out the difficulties faced by some models at low metallicities, like
the absence of GRANADA models for $\log Z/Z_\odot < -0.7$ and the lack
of very metal poor stars in STELIB (see section
\ref{sec:Results_ModXMod}).  On the other hand, these SCs allow us to
test the capability of the models to estimate correct ages even when
$Z$ is a factor of three lower than in the models.

A general conclusion from Fig.\ \ref{fig:Ages_correlations} that our
spectral fitting ages are slightly older than the S-CMD ages. Most of
the points and linear fits are located above the identity line. Only
the GRANADA linear fit, when all the SCs are included, crosses this
line, and exclusively due to the effect of the most metal poor SCs,
whose ages come out severely underestimated due to the lack of SED@
models for $Z < 0.2$ solar. Because stellar evolution runs faster at
higher metallicities, the ages predicted for these metal poor SCs are
much younger if models with isochrones of higher metallicity are used.
In the other sets of models, the ages of these SCs are relatively well
predicted because they include tracks that follow stellar evolution at
metallicities below 0.2 solar ($\log Z/Z_\odot = -1.3$, $-1.7$ and
$-2.3$). This conclusion also holds for the BC fits, despite the
incompleteness of the STELIB library at low $Z$.

This suggests that the capability of the models to match the ages of
metal poor clusters is driven more by the evolutionary tracks than by
the stellar spectral library. This was pointed out before by
Gonz\'alez Delgado et al.\ (1999), who were able to match the ages of
LMC clusters fitting the high order Balmer lines using models with an
empirical library of $Z_\odot$ stars but following the evolution with
tracks at low $Z$. The explanation is simple. The strength and shape
of the Balmer absorption lines depend on the effective temperature and
gravity, but not on $Z$.  The spectral range that we are fitting is
dominated by the Balmer lines and the Balmer jump, which mainly depend
on effective temperature of the main sequence turn off. Thus, the ages
predicted for metal poor clusters that are fitted by models for which
the evolution is described by stellar tracks at higher metallicity
will be smaller than the real age.  As long as the tracks cover the
correct metallicity range, ages come out in good agreement with CMD
data, regardless of the library.

%***FIG***FIG***FIG***FIG***FIG***FIG***FIG***FIG***FIG***FIG***
\begin{figure}
\includegraphics[width=0.5\textwidth]{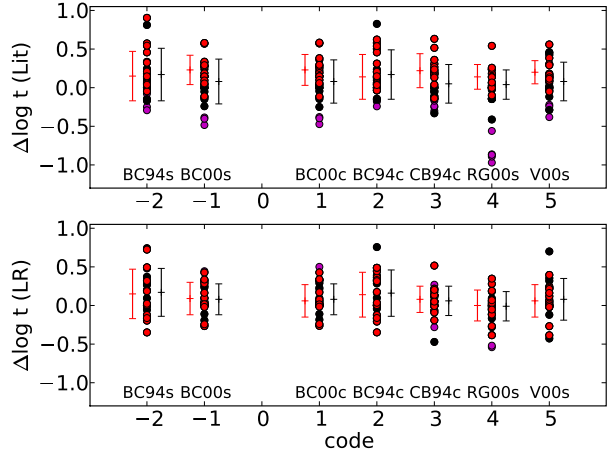}
\caption{Logarithmic difference between the spectral fit age obtained
for each set of models and the literature age tabulated in Table
\ref{tab:Data}.  Magenta circles mark the metal poor clusters ($\log
Z/Z_\odot \leq -1.2$). In the top panel {\sc starlight} ages are
compared to S-CMD ages, while in the bottom panel the {\sc starlight}
ages are compared to the LR03 estimates. Error bars mark the mean and
standard deviation for each set of models considering the whole sample
(in black) and only the 13 clusters in Fig.\ \ref{fig:SpecResidStats}
(red).}
\label{fig:Delta_logt}
\end{figure}
%***FIG***FIG***FIG***FIG***FIG***FIG***FIG***FIG***FIG***FIG***

Fig.\ \ref{fig:Delta_logt} summarizes the comparison of ages.  It
shows the distributions of $\Delta \log t = \log t_{\rm PDF} - \log
t_{\rm Lit}$ for different models.  Red circles mark the 13 SCs
within the nominal range of validity of all models studied here.  In
the top panel the literature age $t_{\rm Lit}$ is that from S-CMD
estimates, while in the bottom one the age obtained from Rose's
spectral indices (LR03) is used. Magenta circles mark the five metal
poorest SCs.  Neglecting these objects and using only the S-CMD ages,
the mean values and sample standard deviation of $\Delta \log t$ are
$0.04 \pm 0.20$ for the RG00s models, $0.12 \pm 0.23$ for CB94c, $0.17
\pm 0.21$ for V00s, and from $0.14 \pm 0.22$ to $0.23 \pm 0.29$ dex
for the STELIB ones.  Somewhat smaller differences are obtained using
the LR03 age estimates: $-0.01 \pm 0.19$ (RG00s), $0.06 \pm 0.19$
(CB94c), $0.08 \pm 0.27$ (V00s), and from $0.08 \pm 0.20$ to $0.17 \pm
0.30$ dex (BC models).  These statistics are represented as error bars
in Fig.\ \ref{fig:Delta_logt}.

In summary, the main conclusions of this section are:

\begin{enumerate}

\item Spectral fits with {\sc starlight} provide ages that are within
a factor better than 2 of the S-CMD ages.

\item The spectral fit ages are more similar to the Rose's age
estimations than to S-CMD ages.

\item Fits with models based on the GRANADA and/or MILES libraries can
date young and intermediate stellar clusters better than the models
with the STELIB library.

\item However, the age of metal poor SCs is not well constrained if
the models do not include evolutionary tracks at the correct
metallicity range.  Thus, at low $Z$, the evolutionary tracks is the
main ingredient in the models to predict the age of young clusters.

\end{enumerate}

\subsection{Metallicities}

In contrast with the age results, the correlation between the {\sc
starlight} metallicities and literature values (from the CaII triplet
or the Rose's indices; see values in Table \ref{tab:Data}) is poor.
This result was expected considering that $\sigma(\log Z/Z_\odot)$ is
typically two times larger than $\sigma(\log t)$, and that $Z$ covers
a 1 dex smaller dynamic range than $t$. This is also a consequence of
the discrete metallicity grids, with a maximum of only $N_Z = 6$ $Z$
values in any given model, covering a wide range from 1/200 to 2.5
solar. Because the step in $Z$ in each set of models is at least of a
factor 2, we consider that the $Z$ is well estimated if the difference
between the {\sc starlight} metallicity and the literature values is
$\le$ a factor 2.\footnote{As discussed in Paper I, interpolating
grids in $Z$ alleviates somewhat the discreteness effects, but the
errors in the interpolated spectra are of the same order of the
spectral residuals obtained in the actual data fits, such that the
overall gain in resorting to interpolated grids is not significant.
Experiments with the V00s models also show that the $Z$-estimates
obtained with $Z$-interpolated grids do not lead to a better agreement
with the literature values.}

As done for the ages, we define $\Delta \log Z/Z_\odot = (\log
Z/Z_\odot)_{\rm PDF} - (\log Z/Z_\odot)_{\rm Lit}$ as the logarithmic
difference between our $Z$ values and those in the literature.
$Z_{\rm Lit}$ can be based either on the CaII triplet or Rose indices
(LR03; see Table \ref{tab:Data}).  These two literature estimates
differ typically by 0.18 dex, with $Z_{\rm LR03}$ smaller than $Z_{\rm
CaT}$. This gives a measure of how consistent these estimates
are. Naturally, this dispersion on our reference values propagates to
$\Delta \log Z/Z_\odot$.

\begin{figure}
\includegraphics[width=0.5\textwidth]{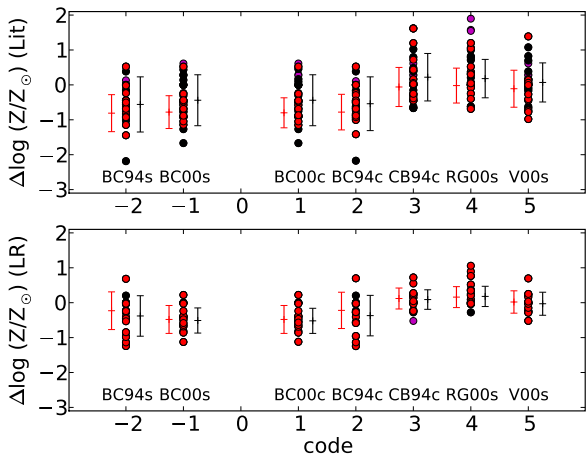}
\caption{As Fig.\ \ref{fig:Delta_logt} but for metallicity.  At the
top panel, the {\sc starlight} $Z$ are compared to CaII triplet
metallicities from Olszewski et al.\ (1991), while in the bottom panel
the comparison is made with respect to the estimates made by LR03.}
\label{fig:Delta_logZ}
\end{figure}

Fig.\ \ref{fig:Delta_logZ} shows the results obtained for each set of
models, in the same format as Fig.\ \ref{fig:Delta_logt}.  The results
that come out from these plots are: 

\begin{enumerate}

\item On average the CaII triplet metallicity differs from our $Z_{\rm
PDF}$ by a factor of 4.

\item Excluding metal poor SCs, only 10$\%$ of the STELIB fits have
metallicities that differ by less than a factor 2 with respect to the
CaII triplet metallicity, but in $\sim$40$\%$ of the MILES or GRANADA
fits the difference is less than a factor 2.

\item STELIB provides metallicities that are significantly
underestimated with respect to both the CaII triplet and LR03
metallicities.  A similar conclusion was derived by Koleva et al.\
(2008) in their analysis of Galactic globular clusters.

\item On average, the Vazdekis models estimate metallicities that
agree better with the CaII triplet metallicity than the estimations
with other models.

\end{enumerate}

%***FIG***FIG***FIG***FIG***FIG***FIG***FIG***FIG***FIG***FIG***
%***FIG***FIG***FIG***FIG***FIG***FIG***FIG***FIG***FIG***FIG***

\subsection{Extinction}

The extinction data for each cluster that are available in the
literature and listed in Table \ref{tab:Data} are very disperse. This
can be seen, for instance, by the lack of correlation between the
extinction derived from the MCPS, the CMD, and McLaughlin \& van der
Marel (2005). These three sets of literature values lead to average
$A_V$ values of 0.41, 0.25 and 0.06, respectively.  Our spectral fits
results are in between them, leading to average extinctions of 0.32,
0.20 and 0.11 for STELIB, MILES and GRANADA models, respectively.
This is as much as can be said about $A_V$ estimates, as the
inhomogeneity of the literature data prevents a more detailed
comparison.

\section{Discussion}
\label{sec:Discussion}

\subsection{Caveats: Stochastic fluctuation effects}

Evolutionary synthesis models always assume that the cluster is
massive enough such that all stages are well sampled, and by applying
these models we have subscribed to this hypothesis.  When a cluster is
not very populated, stochastic fluctuations may play an important role
in the determination of its properties.  Studies dedicated to this
issue have shown that stochastic effects can easily dominate the
integrated light, and that the impact varies with wavelength and age
(e.g. Cervi\~no et al. 2000; Cervi\~no \& Luridiana 2004, 2006;
Lan\c{c}on \& Mouchine 2000). This raises the question of whether we
are entitled to neglect such effects in our analysis.

A way to answer this question is to compared the SC luminosity at a
given wavelength with the contribution of the brightest star in this
band.  We have compiled from the literature the V magnitude of the MC
clusters of the sample.  V ranges from 9.8 to 12.4, with a mean
absolute magnitude of $M_V = -7.2$ or $\log L_V = 34.5$
(erg$\,$s$^{-1}\,$\AA$^{-1}$). The younger SCs of the sample are more
luminous than the mean, and the weakest ones are those with an
intermediate age, 1--2 Gyr.

Following Cervi\~no \& Luridiana (2004) and using the GRANADA models,
we have obtained the minimum luminosity of a SC that ensures that the
fluctuations are less than 10$\%$ of the mean luminosity
($L_{10\%}$). We have also obtained the lowest luminosity limit
($L_{LLL}$), which requires the total luminosity of a cluster to be
larger than the contribution of the brightest star included in the
isochrones. If the luminosity of the cluster is larger than $L_{10\%}$
and $L_{LLL}$, then stochastic effects may be safely neglected.
 
In the V band, these two quantities are $\log L_{10\%} = 34.2$, 33.7,
33.4, and $\log L_{LLL} = 33.4$, 32.7, 32.4
(erg$\,$s$^{-1}\,$\AA$^{-1}$), for ages of 0.2 , 2 and 10 Gyr,
respectively.  These values are well below the average V band
luminosity of our SCs.  Considering that stochastic fluctuations are
even less important in the B and U bands (the spectral range covered
by the spectra analyzed here), we can conclude that these effects are
not important for these clusters. This is true even for the weakest
objects with $\log L_V = 34.0$ (erg$\,$s$^{-1}\,$\AA$^{-1}$) because
they are about 1--2 Gyr old, when the stochastic fluctuations have a
very small impact at the B and U bands (cf.\ Figs.\ 1 and 3 in
Cervi\~no \& Luridiana 2004).

\subsection{Caveats: The abundance ratios}

Another issue that deserves discussion is the inconsistency between
the chemical abundance pattern of the models and data.  In massive
elliptical galaxies, the mismatch between their ``$\alpha$-enhanced''
stellar populations (Worthey, Faber \& Gonzalez 1992) and evolutionary
synthesis models which do not take this into account lead to clearly
identifiable residuals in spectral fits (e.g., Panter et al.\ 2007).
Are analogous effects present in our SCs, and if so, how does this
affect our analysis?

The stellar spectra of the GRANADA models were computed with solar
scaled abundances for all metallicities, while the Galaxev and
Vazdekis models are based on empirical libraries built from nearby
stars, and thus track the relation between $[\alpha/{\rm Fe}]$ and
$[{\rm Fe/H}]$ of the solar neighborhood, where $[\alpha/{\rm Fe}]$
grows from 0 at $Z_\odot$ to $\sim +0.4$ dex at low metallicity (below
$[{\rm Fe/H}] = -0.7$; McWilliam et al.\ 1994). Field stars in the LMC
follow a different pattern (Pomp\'eia et al.\ 2008 and references
therein).  Some $\alpha$-elements like Ca, Si, and Ti show X/Fe
abundance ratios smaller than solar neighborhood stars of the same
metallicity ($[{\rm Fe/H}]$ between $-1.0$ and $-0.5$, approximately),
while others like O and Mg are only slightly deficient. If the LMC
clusters follow the same abundance patterns as the field stars, then
$[{\rm Ca/Fe}] \sim - 0.3$, which is different from both solar-scaled
($[{\rm Ca/Fe}] = 0$) and empirical models at low $Z$ ($[{\rm Ca/Fe}]
\sim + 0.4$).  For O and Mg, however, the empirical libraries should
provide a good match.

In the spectral range that we are fitting, ages are mainly determined
by the Balmer series and break, which are little affected by
variations of $[\alpha/{\rm Fe}]$, and thus the mismatch between the
data and models does not affect significantly our age estimates. The
metallicity, however, could be more affected.  The most $Z$-dependent
features in our spectra are the G band, CaII H$+$K, and Fe lines
(e.g.\ Fe4383, Fe4045).  Because the G band depends mainly on oxygen
and iron abundances, and the LMC [O/Fe] ratio follows the Galactic
distribution, the band should be well fitted by the models. In
contrast, given the difference between data and models in [Ca/Fe], one
would expect that models that fit well the CaII H$+$K lines may fail
to fit the Fe lines, or vice-versa. We have inspected the fit
residuals on the Ca and Fe lines, finding that for most SCs the best
model fits well the two sets of lines. An exception is NGC 1651, which
shows stronger Fe$\lambda 4383$ than any of the models that fit well
the CaII H$+$K lines and G band. 

Overall, we are unable to identify clear signatures of the effect of
the mismatch in abundance pattern between models and our SC data. A
broader spectral coverage (preferably extending to the CaII triplet,
which is strongly affected by the $[{\rm Ca/Fe}]$ ratio; Coelho et
al.\ 2007), and a higher spectral resolution should reveal signs of
this inconsistency. Models and methods which open $Z$ into elemental
abundances have appeared in the recent literature. For instance,
Graves \& Schiavon (2008) present a method to derive elemental
abundances based on Lick indices, while methods to estimate
$[\alpha/{\rm Fe}]$ with full spectral fits have been presented by
Prugniel et al.\ (2007) and Walcher et al.\ (2009), both based on the
Coelho et al.\ (2007) models for $\alpha$-enhanced SSPs. These more
ambitious methodologies are still in their early days, and progress is
expected in the coming years.

\subsection{Age-metallicity relation for LMC clusters}

%***FIG***FIG***FIG***FIG***FIG***FIG***FIG***FIG***FIG***FIG***
 \begin{figure*}
  \centering
  \begin{minipage}{140mm}
   \includegraphics[angle=0,width=\textwidth]{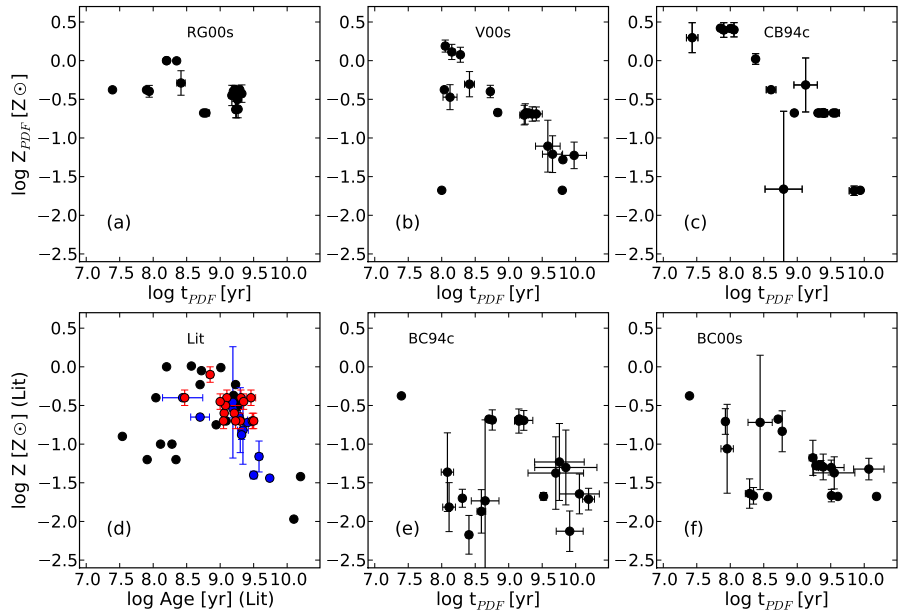}
\caption{Age-metallicity relation for LMC clusters.  
Panels a to c (top, from left to right) and e and f (middle and right bottom), 
show results from {\sc starlight} fits. 
Panel d (bottom left) shows the results from the literature: the
Kerber et al.\ (2007) are plotted in red, LR03 are in blue, and the
results listed in Table \ref{tab:Data} from the S-CMD calibration and
CaII lines are in black.}
\label{fig:t_X_Z_LMC}
 \end{minipage}
\end{figure*}
%***FIG***FIG***FIG***FIG***FIG***FIG***FIG***FIG***FIG***FIG***

In this section we discuss the capability of the models and the method
to reproduce the age-metallicity relation for LMC clusters.  Fig.\
\ref{fig:t_X_Z_LMC}a shows the age-metallicity relation using the data
from the literature, as listed in columns 3 and 4 of Table
\ref{tab:Data}. The plot also includes data for 15 intermediate-age
SCs from Kerber et al.\ (2007), who determine the physical properties
from modelling of HST CMDs. Some of their SCs are in common with this
work. The literature data show a gap in age and in metallicity on the
LMC clusters, in the $3 \leq t \leq 10$ Gyr, and $-0.7 \leq \log
Z/Z_\odot \leq -1.5$ intervals (Olszewski et al. 1991; Girardi et al
1995; MacKey \& Gilmore 2003). So far, only the cluster ESO 121SC-03
was found in the $t$-$Z$ gap (Geisler et al. 1997). This gap has been
interpreted as a consequence of the history of SC formation in the
LMC: two bursts with a significant enrichment in between the two
episodes. This result is reflected in panel a of Fig.\
\ref{fig:t_X_Z_LMC} except for five points corresponding to the young
and intermediate age clusters NGC 1818, NGC 1866, NGC 2133, NGC 2134
and NGC 2214. The metallicities of these clusters are from Sangar \&
Pandei (1989) and Seggeswi \& Ritcher (1989), and these are clearly
underestimated.

Panels b to f of Fig.\ \ref{fig:t_X_Z_LMC} show the $t$-$Z$ relation
obtained here for each set of models. STELIB-based models produce
results that are very inconsistent with the $t$-$Z$ relation. The
metallicity is underestimated for most of the clusters, which, as a
consequence of the $t$-$Z$ degeneracy, move the intermediate-age SCs
towards older ages, filling the age gap.  The problem is even worse
for young clusters, whose estimated metallicities come out severely
underestimated. These conclusions are independent on the evolutionary
tracks and the IMF, and are clearly a consequence of the lack of truly
metal poor stars in the stellar library.  However, as already pointed
out (see Fig.\ \ref{fig:Delta_logt}), the ages are not so badly
predicted.

Fits with the GRANADA models produce metallicities for young and
intermediate-age clusters which are consistent with the average
metallicity $\log Z/Z_\odot = -0.5$ obtained by Kerber et al.\ (2007).
However, because of the lack of predictions for $\log Z/Z_\odot <
-0.7$, these models not only fail to match the metallicity of the old
metal poor SCs, but severely underestimate their ages.

The V00s models produce a very clean correlation between $t$ and $Z$
(with the already explained exception of NGC 1818, the outlier in
Fig.\ \ref{fig:t_X_Z_LMC}e), but the relation indicates a continuous
enrichment and no age gap. The reason is that the clusters NGC 1651,
NGC 1795 and NGC 2121 are $\sim 0.4$ dex younger that the literature
age. It has been argued that Padova 1994 isochrones are better than
the Padova 2000 isochrones (used in the V00s models) because the
latter provide older ages than expected for elliptical galaxies
(BC03), which might explain why these clusters are 0.4--0.5 dex older
with the V00s models than in the literature. The metallicities
derived for these SCs are also incompatible with those in the
literature: The fitted values are 0.7--1 dex smaller than in the
literature.  As usual, a positive difference in $t$ implies a negative
difference in $Z$.

CB94c provides results that are in very good agreement with the LMC
$t$-$Z$ relation. The gap in age and metallicity is reproduced, and
intermediate age clusters have metallicity around $\log Z/Z_\odot =
-0.5$. However, the younger SCs are very metal rich. The consistency
of this result with data in the literature is difficult to evaluate
considering the difficulty to derive metallicity in very young
clusters.

\subsection{Precision on age and metallicity}

We now discuss the precision with which spectral fits using different
models determine the physical properties of a stellar
population. Formal (PDF-based) errors on the determinations of the
age, and metallicity were listed in Tables
\ref{tab:Results_BCBMmc}--\ref{tab:Results_BBC00s}.  As we have
discussed (Figs. 4 and 7) these errors are different for each set of
models, but on average they are 0.07 and 0.13 dex for age and
metallicity, respectively. However, we cannot consider these values as
representative of the accuracy of $t$ and $Z$ estimates from spectral
fits. In fact, because of the large number of degrees of freedom
($N_{dof} \sim 60$), in many cases the formal uncertainty is so
unrealistically small that they are not even listed (the $\pm 0.00$
entries in Tables \ref{tab:Results_BCBMmc}--\ref{tab:Results_BBC00s}).

A first estimate of the accuracy of the $t$ and $Z$ determinations
through spectral fits can be done comparing the results obtained with
different models. For this purpose we have computed the mean $t$ and
$Z$ for each cluster and the dispersion of the different models
results with respect to the global mean. For the whole sample, the
average rms dispersion considering only one of the STELIB models
(BC94c), plus the MILES (CB94c and V00s) and GRANADA (RG00s) models is
0.17 and 0.50 dex for $t$ and $Z$, respectively. Because we have found
that STELIB underestimates $Z$, we have recomputed these values
excluding the STELIB results. In this case, the rms is 0.14 and 0.29
dex for $t$ and $Z$, respectively. We have also tested how this
precision depends on the age of the cluster.  We find an rms of 0.08
dex in $t$ for clusters younger than 1 Gyr, and 0.16 dex for older
ones.  The metallicity, however, has a similar precision for all
ages. This is not unexpected, because for young clusters it is
difficult to constrain $Z$, but the old clusters analyzed here are
mainly metal poor, for which is also difficult to constraint $Z$.

%***FIG***FIG***FIG***FIG***FIG***FIG***FIG***FIG***FIG***FIG***
\begin{figure}
\includegraphics[width=0.5\textwidth]{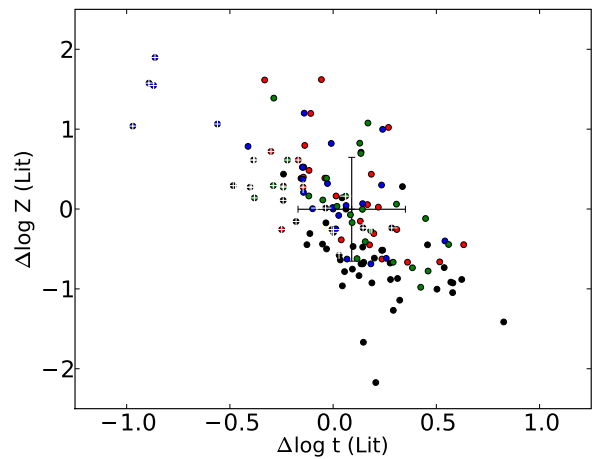}
\caption{Relation between age and metallicity ``errors'', defined as
the logarithmic differences between the spectral fitting and the
literature values.  The cross marks the position of the mean
deviation, and the length of the cross is the rms of the
distributions. STELIB models (only BC00s, and BC94c) results are
plotted in black, CB94c models in red, RG00s models in blue, and V00s
models in green circles. Low metallicity clusters are marked by a
white cross on top the circle.}
\label{fig:dZ_X_dt}
\end{figure}
%***FIG***FIG***FIG***FIG***FIG***FIG***FIG***FIG***FIG***FIG***

We can also estimate the precisions in $t$ and $Z$ from the comparison
between the values estimated with {\sc starlight} and those in the
literature (Table \ref{tab:Data}). In principle, this provides a more
absolute measure than the model-to-model internal variations used
above. The deviations between our $\log t$ and $\log Z/Z_\odot$
estimates and the literature values are plotted against each-other in
Fig.\ \ref{fig:dZ_X_dt}. The rms of these distributions are plotted
with a cross 0.3 dex wide in $\log t$ and 0.7 dex wide in $\log
Z/Z_\odot$. These cannot be considered the precision of the models
because these quantities reflect too the error associated to the
literature values, which can be very uncertain. For example, in Fig.\
\ref{fig:t_X_Z_LMC}a there are five young clusters that are completely
off the age-metallicity relation of the LMC. Moreover, for many
clusters, the ages have been estimated using the S-CMD calibration by
Girardi et al.\ (1995), which by itself introduces 0.137 dex
uncertainty in age. Recently, Pessev et al.\ (2008) have derived a new
S-CMD calibration using the CMD results from Kerber et al.\
(2007). For the clusters that are in common, Kerber et al derive older
ages than Girardi et al. So, the Pessev et al calibration is offset
with respect to the Girardi et al by 0.235 dex. The {\sc starlight}
ages are also older than the literature values, most of which come
from the Girardi et al calibration, but the detected offset towards
older ages is only 0.09 dex on average, lower than the offset found by
Pessev et al. Hence, the literature data do not seem as solid as
desirable to be used as fiducial reference values.

Bearing in mind these caveats, we have computed the statistics of the
difference between our $t$ and $Z$ values, averaged over the CB94c,
RG00s and V00s models, with respect to the literature data. This
results in an rms difference of 0.15 dex in age and 0.36 dex in
metallicity.

In summary, the precision (or consistency between models and
literature data) to constraint the age is $\sim 0.1$--0.2 dex in age
and $\sim 0.3$--0.4 dex in metallicity.

\section{Summary and Conclusions}
\label{sec:Conclusions}

This paper closes our empirical tests of modern spectral models for
stellar populations. This study started in Paper I, where technical
aspects of spectral fitting of SCs and the derivation of age,
metallicity and extinction estimates were presented. In this paper we
applied this methodology to spectra of 27 SCs from LR03 spanning a
wide range of ages and metallicities. These data were fitted with
seven versions of high resolution spectral models presently available:
Galaxev with STELIB and MILES, SED@ with GRANADA, and Vazdekis with
MILES stellar libraries. Extensive comparisons of the quality of the
spectral fits and the inferred physical properties were presented, in
an effort to map the pros and cons of each set of evolutionary
synthesis models and thus provide useful feedback for model makers, as
well as guidance to the ever growing community of users of such
models.

Our main conclusions may be summarized as follows:

\begin{enumerate}

\item Models based on the MILES and GRANADA libraries yield slightly
better spectral fits, as long as their age and metallicity limitations
are observed. Overall, however, all the models are able to produce
very good quality fits.

\item The formal uncertainties in age are on average less than 0.1
dex. The metallicity and extinction PDFs are broader than that of the
age, indicating that $t$ is better constrained than $Z$ and
$A_V$. This is at least in part a consequence of small spectral range
used here to perform the fits. Uncertainties are on average smaller
for MILES and GRANADA models than for STELIB ones.  With the former
two models we obtain $\sigma(\log t) = 0.05$ dex, $\sigma(\log
Z/Z_\odot) = 0.1$ dex, and $\sigma(A_V) = 0.06$ mag, while with the
latter these average values increase to 0.1 dex, 0.2 dex and 0.09 mag,
respectively.

\item Ages correlate very well with the data in the literature.  Ages
derived from our spectral fits are 0.09 dex larger than the S-CMD
ages. 

\item Metallicities derived from spectral fits correlate poorly with
the literature data. This happens due to a combination of the
intrinsic difficulty in deriving $Z$ (specially over the limited
spectral range studied here), the coarseness of the grids in $Z$, plus
the inhomogeneity and uncertainties in the literature values. 

\item Fits with STELIB-based models produce metallicities
systematically smaller by about 0.6 dex with respect to what is found
with other models. Thus, STELIB/BC03 based results will need
revision. In particular, stellar metallicities estimated with BC03
models will probably be revised upwards. However, ages are probably
right.

\item Extinctions derived are small, as expected for the clusters in
our sample.

\item Metal poor clusters are poorly fitted by the GRANADA models as a
consequence of lack of predictions for $Z$ below 0.2 solar.  The error
on the age estimated by the other models is also higher than the
average precision reflecting too the difficulty of derived ages for
metal poor clusters if the Balmer lines are contaminated by the
contribution of blue stars in the horizontal branch.

\item The precision (or consistency) of the models to determine the
age and metallicity is 0.17 and 0.5 dex (rms of the models with
respect to the mean). If STELIB are excluded, the consistency of the
models is better, with an rms of 0.1 and 0.3 dex for age and
metallicity, respectively.

\item Model-to-model dispersions in derived $t$ and $Z$ values are about
  0.2 and 0.5 dex, respectively. Removing models based on STELIB these
  values reduce to 0.1 dex in age and 0.3 dex in metallicity. Similar
  differences are found when comparing the spectral fit results to the
  literature values for $t$ and $Z$.

\end{enumerate}

These conclusions indicate the relevance to
have models with isochrones covering a big range in metallicity and
stellar libraries covering a big range in the stellar parameters
(T$_{eff}$, gravity, and metallicity).

\section*{Acknowledgments}

This work has been funded with support from the Spanish Ministerio de
Educaci\'on y Ciencia through the grants AYA2007-64712, and
co-financed with FEDER funds.  We are grateful to Gustavo Bruzual for
providing the new set of GALAXEV models, Miguel Cervi\~no for the SED@
code, Alejandro Vazdekis for making publicly available his models in
advance of publication, James Rose for kindly sending us the star
cluster spectra, and Jes\'us Ma\'\i z-Apell\'aniz, Jo\~ao Francisco
dos Santos, Miguel Cervi\~no, Claus Leitherer and Eduardo Bica for
discussions.  We also thank the anonymous referee for his/her very
thorough job with both Papers I and II.  We also thank support from a
joint CNPq-CSIC bilateral collaboration grant.  RGD dedicates this
work to her friends Mika and Bene for their lovely taking care of her
during the last year. RGD thanks too to Olga for her friendship, and
Cid's family for their support and hospitality along the years.

%*******************************************************************************************************************
%c     One-line output for tex-tables!
%      write (*,1000) arq_synt_Z(1)(1:5) , 
%     &     BestSSP_logt(ib) , BG_logZ(ib) , BestSSP_A(ib) , 
%     &     BestSSP_rms(ib) , BestSSP_adev(ib) , drc_SmM ,
%     &     mSSP_t_ave(ib) , mSSP_t_sig(ib) ,
%     &     mSSP_rms(ib)   , mSSP_adev(ib)  ,
%     &     D4logt_ave , D4logt_sig , 
%     &     D4logZ_ave , D4logZ_sig , 
%     &     D4A_ave    , D4A_sig    ,
%     &     arq_synt_Z(1)(28:33)

\begin{table*}
\centering
\caption{Results for CB94c models}
\begin{tabular}{@{}lrrrrrrrrrrrrr@{}} \hline        &
\multicolumn{4}{c}{Single population fits} &
\multicolumn{5}{c}{Multi population fits} &
\multicolumn{3}{c}{Bayesian estimates} \\ \hline
Cluster   &
$\log t$  &
$\log Z/Z_\odot$  &
$A_V$     &
$\overline{\Delta}$ &
$\delta_{m}$   & 
$\overline{\log t}_{m}$ &
$\sigma_{m}$  & 
$\overline{\Delta}_{m}$ &
$\log T_{m}$ (\%) &
$\log t$ &
$\log Z/Z_\odot$  & 
$A_V$     \\  
(1)  &
(2)  &
(3)  &
(4)  &
(5)  &
(6)  &
(7)  &
(8)  &
(9)  &
(10) &
(11) &
(12) &
(13) \\ \hline
NGC 411 &  9.26 & -0.68 & 0.17 &  2.4 &  0.09 &  9.26 &  0.26 &  2.3 &  9.26  ($ 34$)  &  9.26 $\pm$  0.00 & -0.68 $\pm$  0.02 & 0.17 $\pm$ 0.04 \\  % 4TeXtable  BCBMmc
NGC 416 &  9.51 & -0.68 & 0.00 &  2.4 &  1.64 &  9.13 &  1.23 &  1.4 &  9.76  ($ 25$)  &  9.53 $\pm$  0.07 & -0.71 $\pm$  0.17 & 0.02 $\pm$ 0.08 \\  % 4TeXtable  BCBMmc
NGC 419 &  9.26 & -0.68 & 0.15 &  2.0 &  0.49 &  9.19 &  0.53 &  1.6 &  8.96  ($ 60$)  &  9.24 $\pm$  0.08 & -0.64 $\pm$  0.17 & 0.15 $\pm$ 0.04 \\  % 4TeXtable  BCBMmc
NGC 1651 &  9.40 & -0.68 & 0.14 &  4.1 &  0.07 &  9.42 &  0.42 &  4.0 &  9.57  ($ 46$)  &  9.40 $\pm$  0.02 & -0.68 $\pm$  0.03 & 0.14 $\pm$ 0.09 \\  % 4TeXtable  BCBMmc
NGC 1754 &  9.94 & -1.68 & 0.29 &  2.5 &  0.11 &  9.75 &  0.96 &  2.3 &  9.95  ($ 50$)  &  9.94 $\pm$  0.00 & -1.68 $\pm$  0.00 & 0.29 $\pm$ 0.05 \\  % 4TeXtable  BCBMmc
NGC 1783 &  8.96 &  0.02 & 0.00 &  2.1 &  1.32 &  8.80 &  0.69 &  1.3 &  8.96  ($ 28$)  &  9.11 $\pm$  0.17 & -0.29 $\pm$  0.35 & 0.14 $\pm$ 0.13 \\  % 4TeXtable  BCBMmc
NGC 1795 &  9.41 & -0.68 & 0.01 &  5.1 &  0.01 &  9.28 &  0.67 &  5.1 &  9.44  ($ 43$)  &  9.41 $\pm$  0.04 & -0.68 $\pm$  0.05 & 0.11 $\pm$ 0.09 \\  % 4TeXtable  BCBMmc
NGC 1806 &  9.32 & -0.68 & 0.42 &  1.9 &  0.24 &  9.31 &  0.47 &  1.8 &  9.54  ($ 34$)  &  9.33 $\pm$  0.01 & -0.68 $\pm$  0.00 & 0.42 $\pm$ 0.04 \\  % 4TeXtable  BCBMmc
NGC 1818 &  7.38 &  0.42 & 0.57 &  1.6 &  0.44 &  7.35 &  0.73 &  1.3 &  6.52  ($ 30$)  &  7.41 $\pm$  0.07 &  0.35 $\pm$  0.16 & 0.48 $\pm$ 0.19 \\  % 4TeXtable  BCBMmc
NGC 1831 &  8.66 & -0.38 & 0.06 &  2.1 &  0.17 &  8.63 &  0.15 &  2.0 &  8.51  ($ 57$)  &  8.61 $\pm$  0.07 & -0.38 $\pm$  0.01 & 0.15 $\pm$ 0.14 \\  % 4TeXtable  BCBMmc
NGC 1846 &  9.30 & -0.68 & 0.24 &  3.0 &  0.11 &  9.27 &  0.49 &  2.8 &  9.01  ($ 42$)  &  9.31 $\pm$  0.00 & -0.68 $\pm$  0.01 & 0.25 $\pm$ 0.07 \\  % 4TeXtable  BCBMmc
NGC 1866 &  8.01 &  0.42 & 0.56 &  1.4 &  0.04 &  8.06 &  0.26 &  1.4 &  8.06  ($ 49$)  &  8.01 $\pm$  0.00 &  0.42 $\pm$  0.03 & 0.56 $\pm$ 0.04 \\  % 4TeXtable  BCBMmc
NGC 1978 &  9.54 & -0.68 & 0.16 &  1.7 &  0.11 &  9.46 &  0.58 &  1.6 &  9.63  ($ 44$)  &  9.54 $\pm$  0.02 & -0.68 $\pm$  0.00 & 0.15 $\pm$ 0.04 \\  % 4TeXtable  BCBMmc
NGC 2010 &  8.06 &  0.42 & 0.10 &  1.8 &  0.04 &  8.09 &  0.17 &  1.7 &  8.06  ($ 61$)  &  8.05 $\pm$  0.03 &  0.41 $\pm$  0.07 & 0.13 $\pm$ 0.08 \\  % 4TeXtable  BCBMmc
NGC 2121 &  9.60 & -0.68 & 0.13 &  4.6 &  0.16 &  9.66 &  1.09 &  4.3 & 10.30  ($ 25$)  &  9.57 $\pm$  0.06 & -0.68 $\pm$  0.01 & 0.17 $\pm$ 0.10 \\  % 4TeXtable  BCBMmc
NGC 2133 &  8.36 &  0.02 & 0.28 &  1.8 &  0.02 &  8.38 &  0.05 &  1.7 &  8.36  ($ 44$)  &  8.38 $\pm$  0.03 &  0.02 $\pm$  0.10 & 0.19 $\pm$ 0.11 \\  % 4TeXtable  BCBMmc
NGC 2134 &  9.01 & -2.28 & 0.00 &  1.5 &  0.08 &  8.96 &  0.18 &  1.4 &  8.96  ($ 46$)  &  8.87 $\pm$  0.22 & -1.94 $\pm$  0.81 & 0.06 $\pm$ 0.07 \\  % 4TeXtable  BCBMmc
NGC 2136 &  7.91 &  0.42 & 0.57 &  1.6 &  0.12 &  7.98 &  0.37 &  1.5 &  7.96  ($ 42$)  &  7.90 $\pm$  0.03 &  0.41 $\pm$  0.07 & 0.60 $\pm$ 0.08 \\  % 4TeXtable  BCBMmc
NGC 2203 &  9.40 & -0.68 & 0.43 &  2.8 &  0.16 &  9.44 &  0.96 &  2.6 &  8.96  ($ 39$)  &  9.39 $\pm$  0.02 & -0.67 $\pm$  0.04 & 0.42 $\pm$ 0.09 \\  % 4TeXtable  BCBMmc
NGC 2210 &  9.83 & -1.68 & 0.21 &  1.5 &  0.45 &  9.90 &  1.14 &  1.2 & 10.28  ($ 21$)  &  9.86 $\pm$  0.09 & -1.69 $\pm$  0.09 & 0.20 $\pm$ 0.05 \\  % 4TeXtable  BCBMmc
NGC 2213 &  9.38 & -0.68 & 0.32 &  2.6 &  0.36 &  9.45 &  0.43 &  2.3 &  8.96  ($ 36$)  &  9.37 $\pm$  0.02 & -0.68 $\pm$  0.02 & 0.32 $\pm$ 0.06 \\  % 4TeXtable  BCBMmc
NGC 2214 &  7.86 &  0.42 & 0.14 &  1.1 &  0.04 &  7.84 &  0.04 &  1.1 &  7.86  ($ 46$)  &  7.85 $\pm$  0.01 &  0.42 $\pm$  0.00 & 0.15 $\pm$ 0.04 \\  % 4TeXtable  BCBMmc
NGC 2249 &  8.96 & -0.68 & 0.00 &  2.1 &  0.18 &  8.81 &  0.46 &  1.9 &  8.96  ($ 80$)  &  8.96 $\pm$  0.00 & -0.68 $\pm$  0.01 & 0.01 $\pm$ 0.02 \\  % 4TeXtable  BCBMmc
47Tuc &  9.99 & -0.38 & 0.00 &  2.9 &  0.43 &  9.87 &  0.88 &  2.5 & 10.01  ($ 24$)  &  9.98 $\pm$  0.03 & -0.37 $\pm$  0.03 & 0.02 $\pm$ 0.02 \\  % 4TeXtable  BCBMmc
M 15 &  9.86 & -1.68 & 0.04 &  1.6 &  1.41 &  9.31 &  1.48 &  1.0 &  9.65  ($ 19$)  &  9.85 $\pm$  0.00 & -1.68 $\pm$  0.01 & 0.05 $\pm$ 0.03 \\  % 4TeXtable  BCBMmc
M 79 &  9.93 & -1.68 & 0.00 &  1.4 &  0.16 &  9.89 &  0.74 &  1.2 &  9.95  ($ 77$)  &  9.93 $\pm$  0.02 & -1.68 $\pm$  0.00 & 0.01 $\pm$ 0.01 \\  % 4TeXtable  BCBMmc
NGC 1851 &  9.63 & -0.68 & 0.00 &  2.6 &  2.42 &  9.47 &  1.04 &  1.5 &  9.76  ($ 69$)  &  9.61 $\pm$  0.05 & -0.68 $\pm$  0.00 & 0.01 $\pm$ 0.01 \\  % 4TeXtable  BCBMmc
\hline
\label{tab:Results_BCBMmc}
\end{tabular}
\end{table*}

\begin{table*}
\centering
\caption{Results for RG00s models}
\begin{tabular}{@{}lrrrrrrrrrrrrr@{}} \hline        &
\multicolumn{4}{c}{Single population fits} &
\multicolumn{5}{c}{Multi population fits} &
\multicolumn{3}{c}{Bayesian estimates} \\ \hline
Cluster   &
$\log t$  &
$\log Z/Z_\odot$  &
$A_V$     &
$\overline{\Delta}$ &
$\delta_{m}$   & 
$\overline{\log t}_{m}$ &
$\sigma_{m}$  & 
$\overline{\Delta}_{m}$ &
$\log T_{m}$ (\%) &
$\log t$ &
$\log Z/Z_\odot$  & 
$A_V$     \\  
(1)  &
(2)  &
(3)  &
(4)  &
(5)  &
(6)  &
(7)  &
(8)  &
(9)  &
(10) &
(11) &
(12) &
(13) \\ \hline
NGC 411 &  9.20 & -0.68 & 0.00 &  2.8 &  0.38 &  8.88 &  0.73 &  2.4 &  9.15  ($ 50$)  &  9.13 $\pm$  0.07 & -0.52 $\pm$  0.17 & 0.06 $\pm$ 0.06 \\  % 4TeXtable  BRG00s
NGC 416 &  9.30 & -0.38 & 0.00 &  4.1 &  3.74 &  8.55 &  1.25 &  1.9 &  6.80  ($ 34$)  &  9.28 $\pm$  0.02 & -0.38 $\pm$  0.00 & 0.01 $\pm$ 0.01 \\  % 4TeXtable  BRG00s
NGC 419 &  9.05 & -0.38 & 0.10 &  2.5 &  0.90 &  8.74 &  0.81 &  1.9 &  8.90  ($ 34$)  &  9.06 $\pm$  0.03 & -0.39 $\pm$  0.07 & 0.09 $\pm$ 0.04 \\  % 4TeXtable  BRG00s
NGC 1651 &  9.20 & -0.38 & 0.00 &  4.6 &  0.15 &  8.89 &  0.86 &  4.3 &  9.35  ($ 60$)  &  9.24 $\pm$  0.04 & -0.50 $\pm$  0.15 & 0.04 $\pm$ 0.04 \\  % 4TeXtable  BRG00s
NGC 1754 &  9.20 & -0.38 & 0.00 &  5.2 &  5.73 &  8.41 &  1.25 &  2.1 &  6.80  ($ 38$)  &  9.22 $\pm$  0.03 & -0.39 $\pm$  0.06 & 0.02 $\pm$ 0.02 \\  % 4TeXtable  BRG00s
NGC 1783 &  9.15 & -0.38 & 0.00 &  2.4 &  0.43 &  8.86 &  0.79 &  2.0 &  9.30  ($ 30$)  &  9.19 $\pm$  0.05 & -0.49 $\pm$  0.15 & 0.03 $\pm$ 0.03 \\  % 4TeXtable  BRG00s
NGC 1795 &  9.30 & -0.68 & 0.00 &  5.6 &  0.21 &  8.90 &  0.98 &  5.4 &  9.15  ($ 62$)  &  9.25 $\pm$  0.04 & -0.53 $\pm$  0.15 & 0.05 $\pm$ 0.05 \\  % 4TeXtable  BRG00s
NGC 1806 &  9.25 & -0.68 & 0.09 &  2.5 &  0.35 &  8.94 &  0.79 &  2.3 &  9.15  ($ 49$)  &  9.25 $\pm$  0.02 & -0.65 $\pm$  0.08 & 0.10 $\pm$ 0.05 \\  % 4TeXtable  BRG00s
NGC 1818 &  7.40 & -0.38 & 0.44 &  1.2 &  0.07 &  7.45 &  0.35 &  1.2 &  7.35  ($ 55$)  &  7.39 $\pm$  0.02 & -0.38 $\pm$  0.03 & 0.45 $\pm$ 0.05 \\  % 4TeXtable  BRG00s
NGC 1831 &  8.75 & -0.68 & 0.01 &  1.5 &  0.13 &  8.76 &  0.14 &  1.4 &  8.80  ($ 73$)  &  8.75 $\pm$  0.01 & -0.68 $\pm$  0.00 & 0.03 $\pm$ 0.03 \\  % 4TeXtable  BRG00s
NGC 1846 &  9.25 & -0.68 & 0.00 &  3.2 &  0.28 &  8.86 &  0.84 &  3.1 &  9.15  ($ 74$)  &  9.24 $\pm$  0.03 & -0.65 $\pm$  0.08 & 0.04 $\pm$ 0.04 \\  % 4TeXtable  BRG00s
NGC 1866 &  8.20 &  0.00 & 0.29 &  1.4 &  0.46 &  8.09 &  0.52 &  1.2 &  8.30  ($ 85$)  &  8.20 $\pm$  0.00 &  0.00 $\pm$  0.00 & 0.29 $\pm$ 0.03 \\  % 4TeXtable  BRG00s
NGC 1978 &  9.30 & -0.38 & 0.00 &  2.6 &  0.47 &  8.94 &  0.91 &  2.3 &  9.35  ($ 76$)  &  9.29 $\pm$  0.02 & -0.38 $\pm$  0.02 & 0.03 $\pm$ 0.03 \\  % 4TeXtable  BRG00s
NGC 2010 &  8.20 &  0.00 & 0.01 &  1.8 &  0.06 &  8.15 &  0.44 &  1.7 &  8.30  ($ 86$)  &  8.20 $\pm$  0.01 &  0.00 $\pm$  0.00 & 0.03 $\pm$ 0.03 \\  % 4TeXtable  BRG00s
NGC 2121 &  9.40 & -0.68 & 0.00 &  5.3 &  0.21 &  9.11 &  1.05 &  4.8 &  9.75  ($ 23$)  &  9.33 $\pm$  0.05 & -0.46 $\pm$  0.14 & 0.11 $\pm$ 0.09 \\  % 4TeXtable  BRG00s
NGC 2133 &  8.35 &  0.00 & 0.26 &  1.7 &  0.11 &  8.40 &  0.15 &  1.6 &  8.35  ($ 46$)  &  8.35 $\pm$  0.01 &  0.00 $\pm$  0.02 & 0.26 $\pm$ 0.05 \\  % 4TeXtable  BRG00s
NGC 2134 &  8.45 & -0.38 & 0.05 &  1.3 &  0.08 &  8.46 &  0.16 &  1.2 &  8.50  ($ 93$)  &  8.39 $\pm$  0.07 & -0.24 $\pm$  0.18 & 0.14 $\pm$ 0.11 \\  % 4TeXtable  BRG00s
NGC 2136 &  7.95 & -0.38 & 0.60 &  1.6 &  0.04 &  7.94 &  0.06 &  1.5 &  7.90  ($ 61$)  &  7.95 $\pm$  0.05 & -0.41 $\pm$  0.10 & 0.62 $\pm$ 0.10 \\  % 4TeXtable  BRG00s
NGC 2203 &  9.30 & -0.68 & 0.11 &  3.0 &  0.27 &  9.03 &  0.81 &  2.7 &  9.15  ($ 42$)  &  9.27 $\pm$  0.04 & -0.55 $\pm$  0.15 & 0.12 $\pm$ 0.07 \\  % 4TeXtable  BRG00s
NGC 2210 &  9.20 & -0.38 & 0.00 &  6.0 & 11.20 &  8.19 &  1.20 &  1.7 &  6.80  ($ 43$)  &  9.20 $\pm$  0.02 & -0.41 $\pm$  0.10 & 0.02 $\pm$ 0.02 \\  % 4TeXtable  BRG00s
NGC 2213 &  9.30 & -0.68 & 0.00 &  2.7 &  0.22 &  9.04 &  0.73 &  2.5 &  9.15  ($ 66$)  &  9.27 $\pm$  0.03 & -0.65 $\pm$  0.08 & 0.04 $\pm$ 0.03 \\  % 4TeXtable  BRG00s
NGC 2214 &  7.90 & -0.38 & 0.15 &  1.2 &  0.00 &  7.90 &  0.00 &  1.2 &  7.90  ($100$)  &  7.90 $\pm$  0.01 & -0.38 $\pm$  0.04 & 0.14 $\pm$ 0.03 \\  % 4TeXtable  BRG00s
NGC 2249 &  8.80 & -0.68 & 0.00 &  2.1 &  0.59 &  8.71 &  0.38 &  1.7 &  8.95  ($ 57$)  &  8.79 $\pm$  0.05 & -0.68 $\pm$  0.00 & 0.06 $\pm$ 0.14 \\  % 4TeXtable  BRG00s
47Tuc &  9.95 & -0.68 & 0.00 &  4.3 &  0.14 &  9.85 &  0.90 &  4.2 & 10.10  ($ 91$)  &  9.90 $\pm$  0.08 & -0.60 $\pm$  0.13 & 0.06 $\pm$ 0.06 \\  % 4TeXtable  BRG00s
M 15 &  9.20 & -0.38 & 0.00 &  7.0 & 12.45 &  7.96 &  1.20 &  1.8 &  6.80  ($ 52$)  &  9.14 $\pm$  0.27 & -0.41 $\pm$  0.14 & 0.06 $\pm$ 0.20 \\  % 4TeXtable  BRG00s
M 79 &  9.20 & -0.38 & 0.00 &  6.9 & 12.62 &  8.40 &  1.21 &  1.9 &  6.80  ($ 36$)  &  9.21 $\pm$  0.03 & -0.43 $\pm$  0.11 & 0.02 $\pm$ 0.02 \\  % 4TeXtable  BRG00s
NGC 1851 &  9.30 & -0.38 & 0.00 &  4.2 &  2.38 &  8.82 &  1.18 &  2.4 &  9.50  ($ 50$)  &  9.31 $\pm$  0.02 & -0.38 $\pm$  0.01 & 0.02 $\pm$ 0.02 \\  % 4TeXtable  BRG00s
\hline
\label{tab:Results_BRG00s}
\end{tabular}
\end{table*}

\begin{table*}
\centering
\caption{Results for BC94c models}
\begin{tabular}{@{}lrrrrrrrrrrrrr@{}} \hline        &
\multicolumn{4}{c}{Single population fits} &
\multicolumn{5}{c}{Multi population fits} &
\multicolumn{3}{c}{Bayesian estimates} \\ \hline
Cluster   &
$\log t$  &
$\log Z/Z_\odot$  &
$A_V$     &
$\overline{\Delta}$ &
$\delta_{m}$   & 
$\overline{\log t}_{m}$ &
$\sigma_{m}$  & 
$\overline{\Delta}_{m}$ &
$\log T_{m}$ (\%) &
$\log t$ &
$\log Z/Z_\odot$  & 
$A_V$     \\  
(1)  &
(2)  &
(3)  &
(4)  &
(5)  &
(6)  &
(7)  &
(8)  &
(9)  &
(10) &
(11) &
(12) &
(13) \\ \hline
NGC 411 &  9.38 & -1.68 & 0.58 &  2.8 &  0.00 &  9.41 &  0.18 &  2.8 &  9.36  ($ 62$)  &  9.38 $\pm$  0.04 & -1.67 $\pm$  0.05 & 0.58 $\pm$ 0.05 \\  % 4TeXtable  BBC94c
NGC 416 & 10.09 & -1.68 & 0.37 &  2.2 &  0.03 & 10.02 &  0.39 &  2.1 & 10.30  ($ 35$)  &  9.98 $\pm$  0.13 & -1.68 $\pm$  0.00 & 0.44 $\pm$ 0.10 \\  % 4TeXtable  BBC94c
NGC 419 &  9.01 & -0.68 & 0.55 &  2.0 &  0.66 &  8.83 &  0.66 &  1.5 &  8.91  ($ 49$)  &  9.18 $\pm$  0.17 & -1.17 $\pm$  0.50 & 0.57 $\pm$ 0.04 \\  % 4TeXtable  BBC94c
NGC 1651 &  9.21 & -0.68 & 0.08 &  4.2 &  0.15 &  8.89 &  0.84 &  3.9 &  9.30  ($ 39$)  &  9.71 $\pm$  0.42 & -1.38 $\pm$  0.47 & 0.47 $\pm$ 0.28 \\  % 4TeXtable  BBC94c
NGC 1754 & 10.29 & -1.68 & 0.00 &  2.6 &  0.17 &  9.80 &  1.16 &  2.4 & 10.30  ($ 66$)  & 10.20 $\pm$  0.09 & -1.75 $\pm$  0.19 & 0.10 $\pm$ 0.10 \\  % 4TeXtable  BBC94c
NGC 1783 &  9.51 & -1.68 & 0.73 &  2.2 &  0.04 &  9.72 &  0.40 &  2.2 &  9.44  ($ 64$)  &  9.52 $\pm$  0.00 & -1.68 $\pm$  0.06 & 0.73 $\pm$ 0.06 \\  % 4TeXtable  BBC94c
NGC 1795 & 10.30 & -1.68 & 0.38 &  5.1 &  0.00 & 10.28 &  0.14 &  5.1 & 10.30  ($ 98$)  & 10.06 $\pm$  0.29 & -1.67 $\pm$  0.28 & 0.51 $\pm$ 0.21 \\  % 4TeXtable  BBC94c
NGC 1806 &  9.16 & -0.68 & 0.31 &  2.1 &  0.59 &  8.89 &  0.67 &  1.7 &  9.28  ($ 38$)  &  9.16 $\pm$  0.01 & -0.68 $\pm$  0.04 & 0.31 $\pm$ 0.05 \\  % 4TeXtable  BBC94c
NGC 1818 &  7.40 & -0.38 & 0.34 &  1.9 &  0.00 &  7.40 &  0.01 &  1.9 &  7.40  ($ 86$)  &  7.40 $\pm$  0.03 & -0.38 $\pm$  0.00 & 0.35 $\pm$ 0.09 \\  % 4TeXtable  BBC94c
NGC 1831 &  8.71 & -0.68 & 0.26 &  2.0 &  0.00 &  8.71 &  0.00 &  2.0 &  8.71  ($100$)  &  8.71 $\pm$  0.02 & -0.68 $\pm$  0.01 & 0.26 $\pm$ 0.05 \\  % 4TeXtable  BBC94c
NGC 1846 &  9.16 & -0.68 & 0.13 &  3.1 &  0.26 &  8.84 &  0.76 &  2.9 &  8.86  ($ 37$)  &  9.15 $\pm$  0.07 & -0.70 $\pm$  0.16 & 0.18 $\pm$ 0.14 \\  % 4TeXtable  BBC94c
NGC 1866 &  8.31 & -1.68 & 0.36 &  2.2 &  0.00 &  8.31 &  0.00 &  2.2 &  8.31  ($100$)  &  8.31 $\pm$  0.04 & -1.72 $\pm$  0.16 & 0.38 $\pm$ 0.10 \\  % 4TeXtable  BBC94c
NGC 1978 &  9.26 & -0.68 & 0.16 &  2.4 &  0.52 &  8.93 &  0.92 &  1.8 &  9.28  ($ 49$)  &  9.86 $\pm$  0.46 & -1.31 $\pm$  0.48 & 0.51 $\pm$ 0.27 \\  % 4TeXtable  BBC94c
NGC 2010 &  8.41 & -2.28 & 0.19 &  2.4 &  0.01 &  8.46 &  0.07 &  2.4 &  8.51  ($ 64$)  &  8.41 $\pm$  0.04 & -2.22 $\pm$  0.21 & 0.13 $\pm$ 0.09 \\  % 4TeXtable  BBC94c
NGC 2121 &  9.34 & -0.68 & 0.16 &  5.2 &  0.26 &  9.08 &  1.26 &  4.5 &  9.30  ($ 29$)  &  9.76 $\pm$  0.37 & -1.24 $\pm$  0.50 & 0.55 $\pm$ 0.35 \\  % 4TeXtable  BBC94c
NGC 2133 &  8.76 & -2.28 & 0.17 &  2.5 &  0.01 &  8.76 &  0.13 &  2.5 &  8.86  ($ 58$)  &  8.66 $\pm$  0.20 & -1.77 $\pm$  1.03 & 0.20 $\pm$ 0.12 \\  % 4TeXtable  BBC94c
NGC 2134 &  8.61 & -1.68 & 0.00 &  1.9 &  0.02 &  8.63 &  0.09 &  1.9 &  8.66  ($ 91$)  &  8.61 $\pm$  0.07 & -1.96 $\pm$  0.30 & 0.10 $\pm$ 0.09 \\  % 4TeXtable  BBC94c
NGC 2136 &  8.11 & -1.68 & 0.51 &  2.3 &  0.02 &  8.22 &  0.34 &  2.3 &  8.21  ($ 61$)  &  8.09 $\pm$  0.10 & -1.40 $\pm$  0.53 & 0.39 $\pm$ 0.13 \\  % 4TeXtable  BBC94c
NGC 2203 &  9.21 & -0.68 & 0.37 &  3.0 &  0.23 &  9.04 &  0.88 &  2.6 &  9.28  ($ 53$)  &  9.23 $\pm$  0.14 & -0.69 $\pm$  0.13 & 0.37 $\pm$ 0.09 \\  % 4TeXtable  BBC94c
NGC 2210 & 10.13 & -2.28 & 0.31 &  2.5 &  0.59 &  8.95 &  1.66 &  2.1 &  6.52  ($ 27$)  &  9.96 $\pm$  0.17 & -2.19 $\pm$  0.21 & 0.33 $\pm$ 0.11 \\  % 4TeXtable  BBC94c
NGC 2213 &  9.16 & -0.68 & 0.22 &  2.3 &  0.19 &  8.98 &  0.70 &  2.1 &  9.28  ($ 57$)  &  9.16 $\pm$  0.00 & -0.68 $\pm$  0.00 & 0.22 $\pm$ 0.05 \\  % 4TeXtable  BBC94c
NGC 2214 &  8.26 & -2.28 & 0.00 &  2.3 &  0.19 &  8.21 &  0.56 &  2.2 &  8.36  ($ 78$)  &  8.15 $\pm$  0.11 & -1.93 $\pm$  0.34 & 0.05 $\pm$ 0.05 \\  % 4TeXtable  BBC94c
NGC 2249 &  8.76 & -0.68 & 0.30 &  2.2 &  0.08 &  8.67 &  0.41 &  2.1 &  8.76  ($ 55$)  &  8.76 $\pm$  0.05 & -0.70 $\pm$  0.18 & 0.30 $\pm$ 0.06 \\  % 4TeXtable  BBC94c
47Tuc &  9.90 & -0.38 & 0.00 &  2.9 &  0.02 &  9.85 &  0.52 &  2.9 &  9.95  ($ 42$)  &  9.86 $\pm$  0.04 & -0.38 $\pm$  0.00 & 0.08 $\pm$ 0.06 \\  % 4TeXtable  BBC94c
M 15 & 10.00 & -2.28 & 0.22 &  2.5 &  1.65 &  8.56 &  1.66 &  1.6 &  8.01  ($ 22$)  &  9.98 $\pm$  0.08 & -2.28 $\pm$  0.01 & 0.22 $\pm$ 0.07 \\  % 4TeXtable  BBC94c
M 79 &  9.68 & -1.68 & 0.00 &  1.9 &  0.17 &  9.73 &  0.64 &  1.7 & 10.30  ($ 16$)  &  9.93 $\pm$  0.28 & -1.94 $\pm$  0.30 & 0.02 $\pm$ 0.02 \\  %%4TeXtable  BBC94c
NGC 1851 &  9.93 & -1.68 & 0.60 &  2.7 &  0.00 &  9.98 &  0.09 &  2.7 &  9.95  ($ 77$)  &  9.96 $\pm$  0.04 & -1.68 $\pm$  0.00 & 0.58 $\pm$ 0.07 \\  % 4TeXtable  BBC94c
\hline
\label{tab:Results_BBC94c}
\end{tabular}
\end{table*}

\begin{table*}
\centering
\caption{Results for BBC00c models}
\begin{tabular}{@{}lrrrrrrrrrrrrr@{}} \hline        &
\multicolumn{4}{c}{Single population fits} &
\multicolumn{5}{c}{Multi population fits} &
\multicolumn{3}{c}{Bayesian estimates} \\ \hline
Cluster   &
$\log t$  &
$\log Z/Z_\odot$  &
$A_V$     &
$\overline{\Delta}$ &
$\delta_{m}$   & 
$\overline{\log t}_{m}$ &
$\sigma_{m}$  & 
$\overline{\Delta}_{m}$ &
$\log T_{m}$ (\%) &
$\log t$ &
$\log Z/Z_\odot$  & 
$A_V$     \\  
(1)  &
(2)  &
(3)  &
(4)  &
(5)  &
(6)  &
(7)  &
(8)  &
(9)  &
(10) &
(11) &
(12) &
(13) \\ \hline
NGC 411 &  9.11 & -1.28 & 0.60 &  2.6 &  0.12 &  9.03 &  0.48 &  2.5 &  8.66  ($ 30$)  &  9.13 $\pm$  0.06 & -1.30 $\pm$  0.09 & 0.59 $\pm$ 0.06 \\  % 4TeXtable  BBC00c
NGC 416 & 10.04 & -1.68 & 0.42 &  2.2 &  0.09 & 10.09 &  0.43 &  2.1 & 10.29  ($ 54$)  & 10.03 $\pm$  0.07 & -1.68 $\pm$  0.00 & 0.44 $\pm$ 0.08 \\  % 4TeXtable  BBC00c
NGC 419 &  9.11 & -1.28 & 0.59 &  1.6 &  0.17 &  9.09 &  0.41 &  1.5 &  9.01  ($ 31$)  &  9.11 $\pm$  0.01 & -1.28 $\pm$  0.01 & 0.58 $\pm$ 0.03 \\  % 4TeXtable  BBC00c
NGC 1651 &  9.36 & -1.28 & 0.36 &  4.0 &  0.13 &  9.18 &  0.74 &  3.9 &  9.57  ($ 43$)  &  9.42 $\pm$  0.17 & -1.31 $\pm$  0.18 & 0.40 $\pm$ 0.16 \\  % 4TeXtable  BBC00c
NGC 1754 & 10.28 & -1.68 & 0.02 &  2.3 &  0.13 & 10.05 &  0.78 &  2.2 & 10.30  ($ 91$)  & 10.24 $\pm$  0.04 & -1.68 $\pm$  0.00 & 0.06 $\pm$ 0.05 \\  % 4TeXtable  BBC00c
NGC 1783 &  9.51 & -1.68 & 0.78 &  2.0 &  0.00 &  9.52 &  0.13 &  2.1 &  9.54  ($ 97$)  &  9.52 $\pm$  0.03 & -1.67 $\pm$  0.07 & 0.78 $\pm$ 0.06 \\  % 4TeXtable  BBC00c
NGC 1795 &  9.36 & -1.28 & 0.31 &  5.0 &  0.07 &  9.01 &  0.98 &  4.9 &  9.54  ($ 33$)  &  9.70 $\pm$  0.37 & -1.45 $\pm$  0.23 & 0.42 $\pm$ 0.20 \\  % 4TeXtable  BBC00c
NGC 1806 &  9.28 & -1.28 & 0.64 &  2.1 &  0.40 &  9.12 &  0.62 &  1.8 &  9.57  ($ 38$)  &  9.28 $\pm$  0.01 & -1.28 $\pm$  0.03 & 0.64 $\pm$ 0.05 \\  % 4TeXtable  BBC00c
NGC 1818 &  7.40 & -0.38 & 0.34 &  1.9 &  0.00 &  7.40 &  0.01 &  1.9 &  7.40  ($ 75$)  &  7.39 $\pm$  0.03 & -0.38 $\pm$  0.00 & 0.35 $\pm$ 0.09 \\  % 4TeXtable  BBC00c
NGC 1831 &  8.71 & -0.68 & 0.32 &  2.0 &  0.00 &  8.72 &  0.02 &  1.9 &  8.71  ($ 79$)  &  8.72 $\pm$  0.03 & -0.68 $\pm$  0.00 & 0.30 $\pm$ 0.06 \\  % 4TeXtable  BBC00c
NGC 1846 &  9.26 & -1.28 & 0.49 &  3.0 &  0.11 &  9.26 &  0.72 &  2.9 &  9.51  ($ 19$)  &  9.24 $\pm$  0.06 & -1.18 $\pm$  0.23 & 0.49 $\pm$ 0.07 \\  % 4TeXtable  BBC00c
NGC 1866 &  8.31 & -1.68 & 0.43 &  2.1 &  0.00 &  8.31 &  0.01 &  2.1 &  8.31  ($ 99$)  &  8.30 $\pm$  0.05 & -1.66 $\pm$  0.12 & 0.43 $\pm$ 0.09 \\  % 4TeXtable  BBC00c
NGC 1978 &  9.48 & -1.28 & 0.49 &  2.2 &  0.31 &  9.21 &  0.85 &  1.8 &  9.57  ($ 37$)  &  9.51 $\pm$  0.18 & -1.30 $\pm$  0.09 & 0.50 $\pm$ 0.07 \\  % 4TeXtable  BBC00c
NGC 2010 &  8.36 & -1.68 & 0.02 &  2.4 &  0.02 &  8.36 &  0.11 &  2.4 &  8.31  ($ 61$)  &  8.35 $\pm$  0.01 & -1.67 $\pm$  0.09 & 0.06 $\pm$ 0.06 \\  % 4TeXtable  BBC00c
NGC 2121 & 10.17 & -1.28 & 0.19 &  4.9 &  0.11 &  9.62 &  1.27 &  4.5 & 10.30  ($ 28$)  & 10.12 $\pm$  0.17 & -1.29 $\pm$  0.07 & 0.24 $\pm$ 0.16 \\  % 4TeXtable  BBC00c
NGC 2133 &  8.31 &  0.00 & 0.27 &  2.4 &  0.00 &  8.30 &  0.06 &  2.4 &  8.31  ($ 75$)  &  8.53 $\pm$  0.18 & -1.11 $\pm$  0.82 & 0.19 $\pm$ 0.11 \\  % 4TeXtable  BBC00c
NGC 2134 &  8.61 & -1.68 & 0.00 &  1.8 &  0.02 &  8.64 &  0.14 &  1.8 &  8.76  ($ 55$)  &  8.56 $\pm$  0.05 & -1.68 $\pm$  0.00 & 0.07 $\pm$ 0.06 \\  % 4TeXtable  BBC00c
NGC 2136 &  7.91 & -0.68 & 0.37 &  2.3 &  0.01 &  7.92 &  0.02 &  2.3 &  7.91  ($ 74$)  &  7.93 $\pm$  0.04 & -0.72 $\pm$  0.20 & 0.35 $\pm$ 0.08 \\  % 4TeXtable  BBC00c
NGC 2203 &  9.36 & -1.28 & 0.65 &  2.8 &  0.12 &  9.51 &  0.78 &  2.6 &  9.54  ($ 20$)  &  9.36 $\pm$  0.02 & -1.27 $\pm$  0.06 & 0.65 $\pm$ 0.07 \\  % 4TeXtable  BBC00c
NGC 2210 &  9.63 & -1.68 & 0.23 &  2.7 &  0.63 &  9.37 &  1.34 &  2.0 & 10.30  ($ 38$)  &  9.62 $\pm$  0.02 & -1.68 $\pm$  0.00 & 0.21 $\pm$ 0.07 \\  % 4TeXtable  BBC00c
NGC 2213 &  9.30 & -1.28 & 0.54 &  2.3 &  0.08 &  9.30 &  0.52 &  2.2 &  9.54  ($ 32$)  &  9.31 $\pm$  0.00 & -1.28 $\pm$  0.01 & 0.54 $\pm$ 0.05 \\  % 4TeXtable  BBC00c
NGC 2214 &  7.86 & -0.68 & 0.00 &  2.1 &  0.04 &  7.83 &  0.38 &  2.1 &  7.91  ($ 88$)  &  7.97 $\pm$  0.09 & -1.20 $\pm$  0.56 & 0.08 $\pm$ 0.07 \\  % 4TeXtable  BBC00c
NGC 2249 &  8.81 & -0.68 & 0.29 &  2.3 &  0.14 &  8.65 &  0.49 &  2.1 &  8.66  ($ 37$)  &  8.78 $\pm$  0.04 & -0.94 $\pm$  0.30 & 0.43 $\pm$ 0.14 \\  % 4TeXtable  BBC00c
47Tuc &  9.48 &  0.00 & 0.00 &  3.1 &  0.09 &  9.43 &  0.61 &  3.0 &  9.60  ($ 33$)  &  9.76 $\pm$  0.15 & -0.30 $\pm$  0.15 & 0.10 $\pm$ 0.07 \\  % 4TeXtable  BBC00c
M 15 &  9.65 & -1.68 & 0.12 &  2.7 &  1.53 &  8.73 &  1.42 &  1.6 &  6.56  ($ 25$)  &  9.64 $\pm$  0.02 & -1.68 $\pm$  0.00 & 0.10 $\pm$ 0.06 \\  % 4TeXtable  BBC00c
M 79 &  9.68 & -1.68 & 0.07 &  1.9 &  0.30 &  9.71 &  1.14 &  1.6 & 10.30  ($ 59$)  &  9.68 $\pm$  0.01 & -1.68 $\pm$  0.00 & 0.07 $\pm$ 0.04 \\  % 
%4TeXtable  BBC00c
NGC 1851 &  9.95 & -1.68 & 0.63 &  2.7 &  0.02 & 10.02 &  0.14 &  2.7 &  9.95  ($ 80$)  &  9.95 $\pm$  0.01 & -1.67 $\pm$  0.05 & 0.62 $\pm$ 0.10 \\  % 4TeXtable  BBC00c
\hline
\label{tab:Results_BBC00c}
\end{tabular}
\end{table*}

\begin{table*}
\centering
\caption{Results for BBC94s models}
\begin{tabular}{@{}lrrrrrrrrrrrrr@{}} \hline        &
\multicolumn{4}{c}{Single population fits} &
\multicolumn{5}{c}{Multi population fits} &
\multicolumn{3}{c}{Bayesian estimates} \\ \hline
Cluster   &
$\log t$  &
$\log Z/Z_\odot$  &
$A_V$     &
$\overline{\Delta}$ &
$\delta_{m}$   & 
$\overline{\log t}_{m}$ &
$\sigma_{m}$  & 
$\overline{\Delta}_{m}$ &
$\log T_{m}$ (\%) &
$\log t$ &
$\log Z/Z_\odot$  & 
$A_V$     \\  
(1)  &
(2)  &
(3)  &
(4)  &
(5)  &
(6)  &
(7)  &
(8)  &
(9)  &
(10) &
(11) &
(12) &
(13) \\ \hline
NGC 411 &  9.38 & -1.68 & 0.57 &  2.8 &  0.00 &  9.39 &  0.14 &  2.8 &  9.36  ($ 75$)  &  9.37 $\pm$  0.04 & -1.67 $\pm$  0.07 & 0.58 $\pm$ 0.05 \\  % 4TeXtable  BBC94s
NGC 416 & 10.11 & -1.68 & 0.30 &  2.1 &  0.05 &  9.99 &  0.67 &  2.0 & 10.30  ($ 42$)  & 10.02 $\pm$  0.13 & -1.68 $\pm$  0.00 & 0.37 $\pm$ 0.10 \\  % 4TeXtable  BBC94s
NGC 419 &  9.01 & -0.68 & 0.55 &  2.0 &  0.57 &  8.89 &  0.45 &  1.5 &  8.91  ($ 36$)  &  9.16 $\pm$  0.17 & -1.14 $\pm$  0.50 & 0.56 $\pm$ 0.04 \\  % 4TeXtable  BBC94s
NGC 1651 &  9.21 & -0.68 & 0.07 &  4.1 &  0.14 &  8.92 &  0.80 &  3.9 &  9.30  ($ 53$)  &  9.54 $\pm$  0.34 & -1.29 $\pm$  0.52 & 0.43 $\pm$ 0.31 \\  % 4TeXtable  BBC94s
NGC 1754 & 10.30 & -2.28 & 0.18 &  2.4 &  0.08 &  9.67 &  1.41 &  2.4 & 10.30  ($ 71$)  & 10.24 $\pm$  0.07 & -2.07 $\pm$  0.29 & 0.15 $\pm$ 0.10 \\  % 4TeXtable  BBC94s
NGC 1783 &  9.51 & -1.68 & 0.71 &  2.2 &  0.03 &  9.67 &  0.38 &  2.2 &  9.44  ($ 48$)  &  9.51 $\pm$  0.00 & -1.68 $\pm$  0.06 & 0.71 $\pm$ 0.05 \\  % 4TeXtable  BBC94s
NGC 1795 & 10.30 & -1.68 & 0.29 &  5.1 &  0.01 & 10.22 &  0.30 &  5.1 & 10.30  ($ 93$)  & 10.05 $\pm$  0.29 & -1.71 $\pm$  0.29 & 0.48 $\pm$ 0.22 \\  % 4TeXtable  BBC94s
NGC 1806 &  9.16 & -0.68 & 0.30 &  2.1 &  0.65 &  8.80 &  0.87 &  1.7 &  9.28  ($ 60$)  &  9.16 $\pm$  0.01 & -0.68 $\pm$  0.04 & 0.30 $\pm$ 0.05 \\  % 4TeXtable  BBC94s
NGC 1818 &  7.40 & -0.38 & 0.34 &  2.0 &  0.00 &  7.40 &  0.00 &  2.0 &  7.40  ($ 95$)  &  7.39 $\pm$  0.03 & -0.38 $\pm$  0.00 & 0.36 $\pm$ 0.09 \\  % 4TeXtable  BBC94s
NGC 1831 &  8.71 & -0.68 & 0.26 &  2.0 &  0.00 &  8.71 &  0.00 &  2.0 &  8.71  ($100$)  &  8.71 $\pm$  0.03 & -0.68 $\pm$  0.01 & 0.25 $\pm$ 0.05 \\  % 4TeXtable  BBC94s
NGC 1846 &  9.16 & -0.68 & 0.12 &  3.1 &  0.30 &  8.80 &  0.85 &  2.9 &  9.28  ($ 36$)  &  9.15 $\pm$  0.08 & -0.72 $\pm$  0.20 & 0.20 $\pm$ 0.16 \\  % 4TeXtable  BBC94s
NGC 1866 &  8.31 & -1.68 & 0.36 &  2.2 &  0.00 &  8.31 &  0.00 &  2.2 &  8.31  ($100$)  &  8.31 $\pm$  0.04 & -1.73 $\pm$  0.17 & 0.37 $\pm$ 0.10 \\  % 4TeXtable  BBC94s
NGC 1978 & 10.30 & -1.68 & 0.53 &  2.3 &  0.01 & 10.28 &  0.09 &  2.3 & 10.30  ($ 92$)  & 10.14 $\pm$  0.31 & -1.57 $\pm$  0.31 & 0.54 $\pm$ 0.15 \\  % 4TeXtable  BBC94s
NGC 2010 &  8.41 & -2.28 & 0.18 &  2.4 &  0.02 &  8.47 &  0.18 &  2.4 &  8.41  ($ 63$)  &  8.42 $\pm$  0.03 & -2.22 $\pm$  0.20 & 0.13 $\pm$ 0.08 \\  % 4TeXtable  BBC94s
NGC 2121 & 10.11 & -1.68 & 0.81 &  5.0 &  0.02 & 10.18 &  0.15 &  5.0 & 10.30  ($ 60$)  &  9.97 $\pm$  0.31 & -1.48 $\pm$  0.40 & 0.65 $\pm$ 0.28 \\  % 4TeXtable  BBC94s
NGC 2133 &  8.76 & -2.28 & 0.17 &  2.5 &  0.01 &  8.77 &  0.13 &  2.4 &  8.86  ($ 54$)  &  8.67 $\pm$  0.20 & -1.81 $\pm$  1.00 & 0.19 $\pm$ 0.11 \\  % 4TeXtable  BBC94s
NGC 2134 &  8.71 & -2.28 & 0.04 &  1.9 &  0.07 &  8.78 &  0.22 &  1.8 &  8.81  ($ 42$)  &  8.62 $\pm$  0.07 & -2.04 $\pm$  0.29 & 0.10 $\pm$ 0.09 \\  % 4TeXtable  BBC94s
NGC 2136 &  8.11 & -1.68 & 0.51 &  2.3 &  0.03 &  8.24 &  0.38 &  2.3 &  8.21  ($ 48$)  &  8.10 $\pm$  0.09 & -1.45 $\pm$  0.49 & 0.39 $\pm$ 0.13 \\  % 4TeXtable  BBC94s
NGC 2203 &  9.21 & -0.68 & 0.37 &  3.0 &  0.22 &  9.07 &  0.79 &  2.6 &  9.30  ($ 38$)  &  9.21 $\pm$  0.09 & -0.68 $\pm$  0.09 & 0.36 $\pm$ 0.07 \\  % 4TeXtable  BBC94s
NGC 2210 &  9.94 & -2.28 & 0.36 &  2.5 &  0.60 &  8.90 &  1.66 &  2.1 &  6.54  ($ 29$)  &  9.88 $\pm$  0.15 & -2.18 $\pm$  0.22 & 0.31 $\pm$ 0.12 \\  % 4TeXtable  BBC94s
NGC 2213 &  9.16 & -0.68 & 0.22 &  2.3 &  0.16 &  9.05 &  0.45 &  2.1 &  9.28  ($ 47$)  &  9.16 $\pm$  0.00 & -0.68 $\pm$  0.00 & 0.22 $\pm$ 0.05 \\  % 4TeXtable  BBC94s
NGC 2214 &  8.26 & -2.28 & 0.00 &  2.3 &  0.20 &  8.21 &  0.54 &  2.2 &  8.36  ($ 83$)  &  8.12 $\pm$  0.10 & -1.86 $\pm$  0.32 & 0.06 $\pm$ 0.05 \\  % 4TeXtable  BBC94s
NGC 2249 &  8.76 & -0.68 & 0.30 &  2.2 &  0.05 &  8.70 &  0.23 &  2.1 &  8.76  ($ 71$)  &  8.76 $\pm$  0.03 & -0.68 $\pm$  0.10 & 0.29 $\pm$ 0.06 \\  % 4TeXtable  BBC94s
47Tuc &  9.89 & -0.38 & 0.00 &  2.9 &  0.02 &  9.84 &  0.52 &  2.9 &  9.99  ($ 23$)  &  9.85 $\pm$  0.04 & -0.38 $\pm$  0.00 & 0.07 $\pm$ 0.05 \\  % 4TeXtable  BBC94s
M 15 &  9.94 & -2.28 & 0.19 &  2.5 &  1.58 &  8.41 &  1.66 &  1.6 &  6.52  ($ 37$)  &  9.94 $\pm$  0.07 & -2.28 $\pm$  0.02 & 0.18 $\pm$ 0.06 \\  % 4TeXtable  BBC94s
M 79 &  9.68 & -1.68 & 0.00 &  1.9 &  0.24 &  9.67 &  0.83 &  1.7 & 10.30  ($ 30$)  &  9.85 $\pm$  0.23 & -1.90 $\pm$  0.29 & 0.02 $\pm$ 0.02 \\  % 4TeXtable  BBC94s
NGC 1851 & 10.03 & -1.68 & 0.46 &  2.6 &  0.01 & 10.02 &  0.12 &  2.6 &  9.98  ($ 50$)  &  9.96 $\pm$  0.05 & -1.68 $\pm$  0.00 & 0.52 $\pm$ 0.08 \\  % 4TeXtable  BBC94s
\hline
\label{tab:Results_BBC94s}
\end{tabular}
\end{table*}

\begin{table*}
\centering
\caption{Results for BBC00s models}
\begin{tabular}{@{}lrrrrrrrrrrrrr@{}} \hline        &
\multicolumn{4}{c}{Single population fits} &
\multicolumn{5}{c}{Multi population fits} &
\multicolumn{3}{c}{Bayesian estimates} \\ \hline
Cluster   &
$\log t$  &
$\log Z/Z_\odot$  &
$A_V$     &
$\overline{\Delta}$ &
$\delta_{m}$   & 
$\overline{\log t}_{m}$ &
$\sigma_{m}$  & 
$\overline{\Delta}_{m}$ &
$\log T_{m}$ (\%) &
$\log t$ &
$\log Z/Z_\odot$  & 
$A_V$     \\  
(1)  &
(2)  &
(3)  &
(4)  &
(5)  &
(6)  &
(7)  &
(8)  &
(9)  &
(10) &
(11) &
(12) &
(13) \\ \hline
NGC 411 &  9.11 & -1.28 & 0.60 &  2.6 &  0.11 &  9.03 &  0.47 &  2.5 &  8.66  ($ 28$)  &  9.13 $\pm$  0.06 & -1.30 $\pm$  0.09 & 0.58 $\pm$ 0.06 \\  % 4TeXtable  BBC00s
NGC 416 & 10.30 & -1.68 & 0.16 &  2.0 &  0.08 & 10.02 &  0.71 &  2.0 & 10.29  ($ 59$)  & 10.13 $\pm$  0.12 & -1.68 $\pm$  0.00 & 0.30 $\pm$ 0.12 \\  % 4TeXtable  BBC00s
NGC 419 &  9.11 & -1.28 & 0.58 &  1.6 &  0.16 &  9.08 &  0.40 &  1.5 &  9.01  ($ 27$)  &  9.11 $\pm$  0.01 & -1.28 $\pm$  0.01 & 0.57 $\pm$ 0.03 \\  % 4TeXtable  BBC00s
NGC 1651 &  9.34 & -1.28 & 0.36 &  4.0 &  0.14 &  9.16 &  0.75 &  3.9 &  9.57  ($ 34$)  &  9.41 $\pm$  0.13 & -1.33 $\pm$  0.19 & 0.41 $\pm$ 0.18 \\  % 4TeXtable  BBC00s
NGC 1754 & 10.22 & -1.68 & 0.00 &  2.3 &  0.21 &  9.87 &  1.13 &  2.1 & 10.30  ($ 49$)  & 10.19 $\pm$  0.04 & -1.68 $\pm$  0.00 & 0.04 $\pm$ 0.04 \\  % 4TeXtable  BBC00s
NGC 1783 &  9.51 & -1.68 & 0.76 &  2.0 &  0.00 &  9.51 &  0.00 &  2.0 &  9.51  ($100$)  &  9.51 $\pm$  0.02 & -1.67 $\pm$  0.06 & 0.76 $\pm$ 0.05 \\  % 4TeXtable  BBC00s
NGC 1795 &  9.36 & -1.28 & 0.29 &  5.0 &  0.06 &  9.01 &  1.01 &  4.9 &  9.54  ($ 38$)  &  9.65 $\pm$  0.35 & -1.44 $\pm$  0.22 & 0.39 $\pm$ 0.20 \\  % 4TeXtable  BBC00s
NGC 1806 &  9.28 & -1.28 & 0.63 &  2.1 &  0.40 &  9.12 &  0.62 &  1.8 &  9.54  ($ 25$)  &  9.28 $\pm$  0.02 & -1.28 $\pm$  0.05 & 0.63 $\pm$ 0.05 \\  % 4TeXtable  BBC00s
NGC 1818 &  7.40 & -0.38 & 0.33 &  2.0 &  0.00 &  7.40 &  0.01 &  2.0 &  7.40  ($ 76$)  &  7.39 $\pm$  0.03 & -0.38 $\pm$  0.00 & 0.35 $\pm$ 0.09 \\  % 4TeXtable  BBC00s
NGC 1831 &  8.71 & -0.68 & 0.31 &  2.0 &  0.00 &  8.71 &  0.02 &  1.9 &  8.71  ($ 90$)  &  8.71 $\pm$  0.02 & -0.68 $\pm$  0.00 & 0.30 $\pm$ 0.06 \\  % 4TeXtable  BBC00s
NGC 1846 &  9.26 & -1.28 & 0.48 &  3.0 &  0.10 &  9.24 &  0.67 &  2.9 &  9.60  ($ 28$)  &  9.23 $\pm$  0.06 & -1.18 $\pm$  0.23 & 0.48 $\pm$ 0.07 \\  % 4TeXtable  BBC00s
NGC 1866 &  8.31 & -1.68 & 0.43 &  2.1 &  0.00 &  8.31 &  0.00 &  2.1 &  8.31  ($100$)  &  8.29 $\pm$  0.05 & -1.66 $\pm$  0.14 & 0.44 $\pm$ 0.09 \\  % 4TeXtable  BBC00s
NGC 1978 &  9.48 & -1.28 & 0.46 &  2.1 &  0.33 &  9.28 &  0.72 &  1.8 &  9.63  ($ 43$)  &  9.55 $\pm$  0.26 & -1.33 $\pm$  0.13 & 0.49 $\pm$ 0.07 \\  % 4TeXtable  BBC00s
NGC 2010 &  8.36 & -1.68 & 0.02 &  2.4 &  0.01 &  8.36 &  0.12 &  2.4 &  8.31  ($ 56$)  &  8.35 $\pm$  0.02 & -1.67 $\pm$  0.09 & 0.06 $\pm$ 0.06 \\  % 4TeXtable  BBC00s
NGC 2121 & 10.17 & -1.28 & 0.11 &  4.9 &  0.14 &  9.54 &  1.33 &  4.5 & 10.29  ($ 19$)  & 10.09 $\pm$  0.22 & -1.36 $\pm$  0.17 & 0.31 $\pm$ 0.25 \\  % 4TeXtable  BBC00s
NGC 2133 &  8.31 &  0.00 & 0.27 &  2.4 &  0.00 &  8.30 &  0.04 &  2.4 &  8.31  ($ 97$)  &  8.48 $\pm$  0.19 & -0.89 $\pm$  0.88 & 0.20 $\pm$ 0.11 \\  % 4TeXtable  BBC00s
NGC 2134 &  8.61 & -1.68 & 0.00 &  1.8 &  0.02 &  8.64 &  0.16 &  1.8 &  8.76  ($ 63$)  &  8.56 $\pm$  0.05 & -1.68 $\pm$  0.00 & 0.07 $\pm$ 0.06 \\  % 4TeXtable  BBC00s
NGC 2136 &  7.91 & -0.68 & 0.37 &  2.3 &  0.01 &  7.92 &  0.02 &  2.3 &  7.91  ($ 72$)  &  7.93 $\pm$  0.05 & -0.73 $\pm$  0.22 & 0.34 $\pm$ 0.09 \\  % 4TeXtable  BBC00s
NGC 2203 &  9.34 & -1.28 & 0.65 &  2.8 &  0.12 &  9.46 &  0.76 &  2.6 &  9.57  ($ 20$)  &  9.35 $\pm$  0.00 & -1.27 $\pm$  0.06 & 0.64 $\pm$ 0.07 \\  % 4TeXtable  BBC00s
NGC 2210 &  9.60 & -1.68 & 0.16 &  2.7 &  0.63 &  9.43 &  1.16 &  2.0 & 10.30  ($ 23$)  &  9.61 $\pm$  0.02 & -1.68 $\pm$  0.00 & 0.17 $\pm$ 0.07 \\  % 4TeXtable  BBC00s
NGC 2213 &  9.30 & -1.28 & 0.53 &  2.3 &  0.08 &  9.30 &  0.48 &  2.2 &  9.16  ($ 26$)  &  9.30 $\pm$  0.00 & -1.28 $\pm$  0.01 & 0.53 $\pm$ 0.05 \\  % 4TeXtable  BBC00s
NGC 2214 &  7.86 & -0.68 & 0.00 &  2.1 &  0.04 &  7.82 &  0.42 &  2.1 &  7.91  ($ 84$)  &  7.98 $\pm$  0.09 & -1.22 $\pm$  0.56 & 0.08 $\pm$ 0.07 \\  % 4TeXtable  BBC00s
NGC 2249 &  8.81 & -0.68 & 0.28 &  2.3 &  0.13 &  8.67 &  0.39 &  2.1 &  8.66  ($ 35$)  &  8.78 $\pm$  0.04 & -0.83 $\pm$  0.27 & 0.38 $\pm$ 0.12 \\  % 4TeXtable  BBC00s
47Tuc &  9.48 &  0.00 & 0.00 &  3.1 &  0.09 &  9.42 &  0.66 &  3.0 &  9.54  ($ 44$)  &  9.77 $\pm$  0.13 & -0.32 $\pm$  0.14 & 0.09 $\pm$ 0.07 \\  % 4TeXtable  BBC00s
M 15 &  9.63 & -1.68 & 0.04 &  2.8 &  1.55 &  8.73 &  1.38 &  1.6 &  6.56  ($ 23$)  &  9.64 $\pm$  0.02 & -1.68 $\pm$  0.00 & 0.07 $\pm$ 0.05 \\  % 4TeXtable  BBC00s
M 79 &  9.68 & -1.68 & 0.04 &  1.9 &  0.25 &  9.78 &  0.73 &  1.6 & 10.30  ($ 30$)  &  9.67 $\pm$  0.01 & -1.68 $\pm$  0.00 & 0.04 $\pm$ 0.03 \\  % 4TeXtable  BBC00s
NGC 1851 &  9.98 & -1.68 & 0.55 &  2.6 &  0.04 & 10.05 &  0.16 &  2.6 &  9.95  ($ 71$)  &  9.96 $\pm$  0.00 & -1.68 $\pm$  0.02 & 0.56 $\pm$ 0.07 \\  % 4TeXtable  BBC00s
\hline
\label{tab:Results_BBC00s}
\end{tabular}
\end{table*}
%*******************************************************************************************************************

\end{document}